\documentclass[12pt,preprint]{aastex}

\newcommand{\begit}{\begin{itemize}}
\newcommand{\enit}{\end{itemize}}
\newcommand{\begen}{\begin{enumerate}}
\newcommand{\enen}{\end{enumerate}}

\newcommand       \be           {\begin{equation}}
\newcommand       \ee           {\end{equation}}
\newcommand       \bea          {\begin{eqnarray}}
\newcommand       \eea          {\end{eqnarray}}

\newcommand       \keV		{\,{\rm keV }}
\newcommand       \kms		{\,{\rm km \,\, s}^{-1}}
\newcommand       \cm		{\,{\rm cm }}
\newcommand       \pc		{\,{\rm pc }}
\newcommand       \yr		{\,{\rm yr }}
\newcommand       \s		{\,{\rm s }}
\newcommand       \dynes	{\,{\rm dynes }}
\newcommand       \g		{\,{\rm g }}
\newcommand       \kpc		{\,{\rm kpc }}
\newcommand       \K		{\,{\rm K }}

\newcommand       \erg		{\,{\rm erg }}
\newcommand       \ergs		{\,{\rm erg \,\, s}^{-1}}

\newcommand{\beqa}{\begin{eqnarray}} 
\newcommand{\eeqa}{\end{eqnarray}}

\shorttitle{The Carina Nebula Bubble}
\shortauthors{Harper-Clark and Murray}

\begin{document}

\title{One Dimensional Dynamical Models of the Carina Nebula Bubble}
\author{E. Harper-Clark\altaffilmark{1} and N. Murray\altaffilmark{1,2}}
\altaffiltext{1}{Canadian Institute for Theoretical Astrophysics, 60
St.~George Street, University of Toronto, Toronto, ON M5S 3H8, Canada} 
\altaffiltext{2}{Canada Research Chair in Astrophysics}
\email{h-clark@astro.utoronto.ca, murray@cita.utoronto.ca}

\begin{abstract}


We have tested the two main theoretical models of bubbles around massive star 
clusters, Castor et al.~and Chevalier \& Clegg, against observations of the 
well studied Carina Nebula.  The Castor et al.~theory over-predicts the X-ray 
luminosity in the Carina bubble by a factor of 60 and expands too rapidly, by 
a factor of 4; if the correct radius and age are used, the predicted X-ray 
luminosity is even larger.  In contrast, the Chevalier \& Clegg model 
under-predicts the X-ray luminosity by a factor of 10.  We modify the Castor et 
al.~theory to take into account lower stellar wind mass loss rates, radiation 
pressure, gravity, and escape of or energy loss from the hot shocked gas.  We 
argue that energy is advected rather than radiated from the bubble. We 
undertake a parameter study for reduced stellar mass loss rates and for various 
leakage rates and are able to find viable models.  The X-ray surface brightness 
in Carina is highest close to the bubble wall, which is consistent with 
conductive evaporation from cold clouds. The picture that emerges is one in 
which the hot gas pressure is far below that found by dividing the 
time-integrated wind luminosity by the bubble volume; rather, the pressure in 
the hot gas is set by pressure equilibrium with the photoionized gas at 
$T=10^4\K$. It follows that the shocked stellar winds are not dynamically 
important in forming the bubbles.

\end{abstract}


\section{INTRODUCTION}

Over a dynamical time only 2\% of the gas in a typical disk galaxy is
turned into stars \citep{kennicutt98}, in spite of the fact that the
disks are marginally gravitationally stable \citep{kennicutt89}.  This
puzzling result suggests that something other than gravity plays a
(negative) role in star formation. Passive support against collapse
and subsequent star formation can come from magnetic fields or
turbulence. Examination of star forming galaxies and individual star
forming regions commonly shows evidence for more active support
against gravity, for example, expanding bubbles. Active mechanisms,
which come under the rubric of `feedback', include energy or momentum
input by active galactic nuclei, or from stars. Since most disk
galaxies, including our own Milky Way, show little or no nuclear
activity, and since bubbles in individual star forming regions are
clearly not due to active galactic nuclei, in this paper we focus on
feedback from stars. 

Stars of type B2 and earlier supply several types of feedback,
including protostellar jets, main-sequence stellar winds, radiation
pressure, gas pressure associated with ionizing radiation, and
supernovae. The mechanical energy imparted via a supernova is
comparable to that injected by the star's stellar wind over its
lifetime \citep{cas75}. However, it is clear that many
observed bubbles have formed before any of the stars in the central
cluster have exploded in a supernova. Since we are interested in
bubble formation, we will ignore supernovae, leaving open the question
of whether supernovae provide the feedback necessary to limit the rate
of star formation.

There are two competing theoretical models of stellar wind bubbles in
the literature: that of Castor et al.~(i.e.~\citet{cas75} and
\citet{wea77}) and that of \citet{che85}, as implemented in, e.g.,
\citet{ste03}.  In the theory explored by Castor, the stellar-wind
shocked gas is confined by a cool shell of swept up interstellar
medium (ISM).  In contrast, the Chevalier \& Clegg theory ignores
any surrounding material and simply has a steady-state wind.  The
temperatures, pressures and predicted sizes of the X-ray emitting
regions in the two theories are significantly different.  

However, these differences can be masked if the stellar content of the
bubble is not well constrained. Recent work by \citet{smi06a} provides a
detailed accounting of the number of early type stars in the well
studied Carina nebula. Combined with the wealth of multi-wavelength
observations available for Carina, this allows for a stringent test of
the models.

Observations of star forming regions in the Milky Way show that the
gas density is highly variable. In particular, the projected surface
density is consistent with a log-normal distribution, e.g.,
\citet{goodman,wong}. These observations are consistent with both
analytic and numerical studies of supersonic turbulent flows, which
predict log-normal density and column density distributions
\citep{passot,osg01}. These results call into question one of the
fundamental assumptions of the Castor et al.~model, that the gas
surrounding the central star or star cluster has a rather uniform
density distribution. If the mass distribution is non-uniform, the
shell swept up by the shocked stellar wind is likely to have holes. If
so, this will lead to incomplete confinement of the hot gas and a
consequent reduction in the pressure and associated X-ray emission of
the hot gas. Such a situation would be intermediate between the Castor
et al.~model of a completely confined wind and the Chevalier \& Clegg
picture of a free-flowing wind. We explore such intermediate models
below.

This paper is organized as follows. In the next section we summarize
some relevant observations of Carina, as well as some order of
magnitude estimates of the pressures and forces acting on the
surrounding cold, dense gas. In \S \ref{sec:methods} (and the
appendices) we describe the Chevalier \& Clegg model, the Castor et
al. model, and our modifications to the Castor et al.~models. We also
describe our numerical methods. In \S \ref{sec:results} we present the
results of our modeling, including a parameter study of the (modified)
Castor et al.~bubbles. In \S \ref{sec:globules} we discuss the X-ray emission.
In \S \ref{sec:discussion} we discuss the
results and compare to previous work. We give our conclusions in the
final section.

\section{OBSERVATIONS OF CARINA \label{sec:Observations}}

The distance to Carina is $2.3 \,\kpc$ \citep{ah,smi06b}. The complex
contains about 70 O stars \citep{smi06a}, of which 47 (including
evolved massive stars) reside in Trumpler 16 (Tr 16), a star cluster
with a radius $r\sim3\pc$. This makes Carina one of the most massive
(and best studied) star formation regions in the Milky Way. The
bolometric luminosity emerging from the stars is
$9.6\times10^{40}\ergs$. This is about twice the far infrared
luminosity of the complex \citep{smb07}, indicating that half the
stellar flux is intercepted by dust grains within $\sim10-20\pc$ of
the stars. 

We note that this rather low ratio of infrared to bolometric
luminosity is consistent with the fact that we can see Tr 16 in the
optical, and with turbulent models of the ISM, but it is inconsistent
with the assumption of the Castor et al.~model that the ISM and
swept-up bubble shell surrounding the cluster is homogeneous. It
suggests that hot gas can escape the bubble, a point we return to later.

The number of ionizing photons emitted per second is
$Q=9\times10^{50}\s^{-1}$. The kinetic energy of the stellar winds is
$L_w\sim3.5\times10^{38}\ergs$, assuming standard estimates of the mass
loss rates; we argue below that this is likely to be an overestimate
by a factor of $3-10$. The massive stars of Tr 16 are responsible for
about $70\%$ of the bolometric and stellar wind luminosity and of the
ionizing flux of the entire nebula; see Table~\ref{carina}.

A reasonably reliable estimate for the age of Tr 16 is given by the
age of $\eta\ $ Carina, currently the most massive star in the region. $\eta\ $
Carina is $\sim3.6$ million years old, since it has clearly evolved
off the main sequence, while estimates of its initial mass based on
its current luminosity and stellar models \citep{bressan} are in the
range of $120M_\odot$. We thus assume the cluster is approximately 3.6
million years old. It is believed that no supernovae have occurred in
Tr 16 \citep{ret83,lop84,whi94}, but in any case the integrated energy
deposited by the winds, $4\times10^{52}\erg$, substantially exceeds
the energy input of a supernova, of order $10^{51}\erg$.  

For the 17th anniversary of the Hubble Space Telescope, \citet{hubca}
took a 50 light-year-wide mosaic of the centre of the Carina nebula.
This image has unprecedented detail of globules within the bubble and
star birth at the edge of the bubble.  Triggered star formation around
the edge of the bubble in Carina is also observed in X-ray and Near-IR
studies of Carina \citep{san07}.  There are many directional markers,
such as pillars, showing that the bubble edge is dominated by the
effects of Tr 16.  The images show that the main bubble shell is
markedly non-spherical. Smaller bubble structures within the main
bubble can be seen.  \citet{smb07} compiled a multi-wavelength
analysis of the Carina nebula enabling approximate edges to the bubble
to be located. We estimate radii in the range $10-20\pc$.

There have been many observations of Carina in X-rays, which establish
the existence of diffuse emission from the nebula itself
\citep{sew79,fla05,ham07,ham07b}.  The largest area coverage of the
Carina region is provided by \citet{sew79}, who used the Einstein
Observatory to find a diffuse X-ray emission of $4.8 \times
10^{34}$erg s$^{-1}$ for the $0.5-3.0 \keV$ band (scaled from their distance
estimate of $2.6\kpc$ to $2.3\kpc$).

The pressure inside the bubble surrounding Tr 16 contributes directly
to the outward force on the surrounding material. We estimate the
pressure in several ways. 

First, the pressure can be calculated from the free-free radio
emission, e.g., as reported in \citet{huc75}, who
measured a flux of $S_{\nu} = 1.43 \times 10^3$ Jy.  From calculations
of the expected Bremsstrahlung radio flux at the observer
\be  \label{eqn:free-free} 
n_{HII} = 3.6 \left( \frac{R}{pc} \right)^{-3/2}
\left( \frac{D}{kpc} \right)   
\left( \frac{T}{K} \right)^{1/4}   \left( \frac{S_{\nu}}{Jy}
\right)^{1/2}
\epsilon_{HII}^{-1/2}\cm^{-3} ,
\ee 
where $\epsilon_{HII}$ is the volume filling factor of the H II gas
inside the bubble.  For the radius of the bubble, $R = 10 \pc$ and distance, $D = 2.3\kpc$, we find
$n_{HII} = 100\cm^{-3}$ for $T_{HII} = 10^4$K.  Thus, $P = 1.4 \times
10^{-10}\epsilon_{HII}^{-1/2}\dynes\, \cm^{-2}$.

As a consistency check, we note that pressure of an H II region powered
by a cluster emitting $Q$ ionizing photons per second is given by
\be  \label{eqn:Q} 
P_{HII}=\sqrt{3Q\over 4\pi r^3\alpha_r \epsilon_{HII}}k_b T_{HII}
\ee  
where $\alpha_r \approx4\times10^{-13}\cm^3\s^{-1}$ is the recombination
coefficient, and $T_{HII}\approx10^4\K$. Using
$Q=9 \times 10^{50}\s^{-1}$ and $r\approx10\pc$, we find
$P_{HII}\approx 2\times10^{-10}\epsilon_{HII}^{-1/2}\dynes\cm^{-2} $. 

Second, we can use the observed X-ray flux to estimate the pressure in
the hot ($T_x \sim10^7\K$) gas. From Figure 1 in \citet{sew79} we
estimate the radius of the diffuse X-ray emission region to be $\sim
12\pc$, similar to the size of the free-free emission region. Combined
with the diffuse X-ray luminosity and $T_x =6\times10^6\K$, this yields
$n_x\approx9\times10^{-2}$ and
$P_x\approx7\times10^{-11}\dynes\cm^{-2}$, which we take to be
equal to the pressure of the radio emitting gas.

The surface brightness of both the radio and X-ray emission is not
uniform, indicating that the pressure is not constant inside the
bubble. For example, for the two classical and smaller scale embedded
H II regions Carina I and Carina II, using the fluxes and sizes from
\citet{huc75} we find a pressure
$9\times10^{-10}\dynes\cm^{-2}$. Similarly, \citet{sew79} estimate the
density of the hot X-ray emitting gas in the prominent $r\sim1.25\pc$
diffuse patches to be $n\sim0.4$, giving
$P\approx3\times10^{-10}\dynes\cm^{-2}$.

Finally, we can estimate the pressure from the outflow seen to be
emerging from the globule known as `the finger', which has a dense
ionization front and photoevaporative flow, with a terminal shock
resolved close to the edge.  Estimates of the ionization front density
can be obtained from the S II emission line, and from the H$\alpha$
emission measure.  These give electron densities of 2000-6000
cm$^{-3}$ at various positions.  The dense ionization front drives a
photoevaporative flow out into the H II region until it reaches a
terminal shock where the pressure is balanced with the ambient medium.
To calculate $P_{IF}$ assume the temperature is $10^4\K$ due to
hydrogen ionization and the flow is spherical.  Nathan Smith (private
communication) made two estimates, one using the tip of the finger
with a very small ionization front radius but a relatively high number
density of $6000\cm^{-3}$ and a second using the main part of the fist
with a larger ionization front radius and a lower density of
$3300\cm^{-3}$.  These yield $5.4 \times 10^{-10}$dyne cm$^{-2}$ and
$7.3 \times 10^{-10}$dyne cm$^{-2}$, respectively.

These rather direct measurements of the pressure in several gas phases
can be compared to two other pressures, the pressure exerted by the
self-gravity of the surrounding molecular gas (as estimated from CO
measurements) and the pressure of radiation from the stars in Tr 16 on
dust grains in the interstellar medium. \citet{yonekura} find a mass
of molecular gas $M_g\approx3.5\times10^5M_\odot$ in a region of
radius $r\approx20\pc$, yielding a surface density
$\Sigma\approx6\times10^{-2}\g\cm^{-2}$. Assuming this gas is in
hydrostatic equilibrium (there is evidence that it is expanding
outward, but we ignore this for the moment) the dynamical pressure is
$P\approx\pi G\Sigma^2\approx8\times10^{-10}\dynes\cm^{-2}$.

The near equality between the pressure in the X-ray gas and that of
the H II gas likely reflects the fact that the sound crossing time (of
the cooler gas) is similar to the dynamical time or the age of the
system. Like the hot gas, the H II gas will `head for the exits';
pressure gradients will drive the warm gas toward any holes in the
bubble shell, reducing the pressure on a (H II gas) sound crossing
time. 

Unlike the hot gas, the velocity of the warm gas is comparable
to the velocity of the bubble wall, so a pressure driven flow will
not reduce the H II pressure in the bubble by a large (factor of 10 or
more) amount. Thus it appears natural that the H II gas will have a
pressure similar to that estimated from the ionizing photon
luminosity. Conversely, the hot gas pressure will, if it ever reaches
values similar to $L_w\tau/V$, drive outflows through holes in the
bubble that have velocities of order the sound speed of the hot gas,
which is two orders of magnitude larger than the velocity of the
bubble wall. This will lower the pressure of the hot gas until it
approaches that of the H II gas, at which point the pressure gradients
in the two components are similar.

Using the dynamical argument that the H II pressure and the X-ray gas
pressure should be roughly equal, we can draw one more conclusion:
that $\epsilon_{HII}$ is of order unity. To do so, set the right hand
side of eqn. \ref{eqn:Q} equal to the (observed) pressure of the x-ray
gas and solve for $\epsilon_{HII}$, then use the known values of $Q$,
$r$, and $T_{HII}$.

The pressure exerted by radiation is $P_{rad}=L_{bol}/(4\pi r^2 c)\approx
4\times10^{-10}(10\pc/r)^2\dynes\cm^{-2}$.  This near-equality, unlike
that between the H II and X-ray gas pressure, is a bit of a
coincidence. The radiation pressure scales as $R^{-2}$ while the H II
gas pressure scales as $R^{-3/2}$; the radiation pressure was more
important in the past, while the gas pressure will become more
important as the bubble expands in the future.

All of these observationally driven estimates are well below the
pressure estimated by dividing the total stellar wind kinetic energy
by the volume of the bubble; the latter is
$P_{wind}=3.6\times10^{-7}(10\pc/r)^3(L_w/4\times10^{38}\ergs)\dynes\cm^{-2}$.
Either the wind has expanded over a volume larger by a factor of
$\sim1000$, the wind luminosity is overestimated by a similar factor,
the shocked gas loses most of its internal energy (but not by
radiation in the X-ray band), or some combination of all three. We
note that if the hot gas expanded at its sound speed, it would reach
radii of $1\kpc$, so the notion that the hot gas has expanded through
holes in the ISM to occupy region with a typical size of $100\pc$ is
not implausible.

Assuming an age of $3.6$ Myrs, we find an average expansion velocity
for the bubble wall of $\sim\pm3 \,(R/10\pc)\kms$, similar to the $5\kms$
splitting seen in CO \citep{grabelsky}. Observations of ionized gas
typically find larger velocity splittings, of order $\pm20\kms$,
ranging up to several hundred kilometers per second
e.g.~\citet{wh75,wh82}, but we regard this as a measure of the
velocity of gas in the interior of the bubble, not a measurement of
the velocity of the bubble wall.

To sum up, the warm and hot gas appear to be in pressure equilibrium,
with a pressure $P_g\approx2\times10^{-10}\dynes\cm^{-2}$. This is
similar to the radiation pressure, and to the pressure exerted by the 
surrounding, self-gravitating molecular cloud. The kinetic energy
input from stellar winds greatly exceeds the integral of this pressure
over the apparent size of the H II region, $R\approx10\pc$,
suggesting that the winds have escaped to occupy a much larger volume,
or been otherwise dissipated.

\section{METHODS \label{sec:methods}}

\subsection{Chevalier and Clegg bubble}

\citet{ste03} applied the \citet{che85} steady state wind to super
star clusters.  The flow is assumed to pass through the sonic point at
the cluster edge where the source terms become zero.  The equations
describing the flow are given in Appendix I.

Numerical integration of the flow (equations (\ref{cc1}) to
(\ref{cc4})) using iterative calculations of the mach number (equation
(\ref{mach1}) or (\ref{mach2})) at every step enables the flow to be
accurately calculated.  The X-ray luminosity from Bremsstrahlung is
numerically integrated from the resulting temperature and density
profiles. The actual X-ray luminosity will be somewhat larger than
this estimate by about a factor of three at $T_x =10^7\K$, see, e.g.,
\citet{sut93}.

\subsection{Castor bubble}

\begin{figure}
\begin{center}
\epsscale{.90}
\plotone{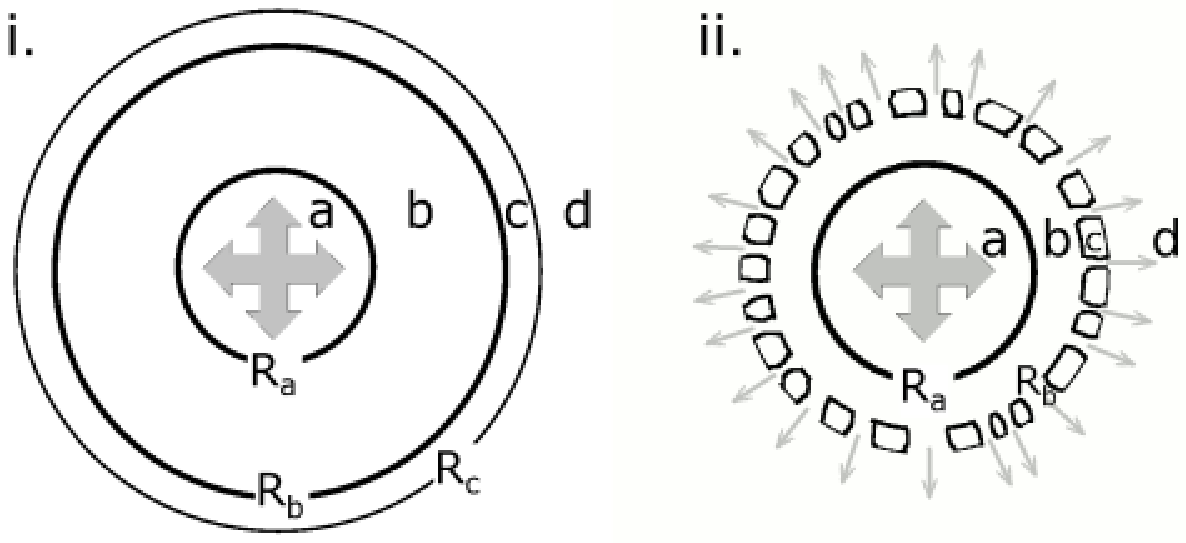}
\caption{{\small Cartoons of the Castor et al.~bubble (i.) and the modified Castor et al.~bubble (ii.).
    The Castor et al.~bubbles have a four zone structure. 
    Zone a is hypersonic stellar wind; zone b is shocked
    stellar wind and evaporated material from the shell; zone c is the
    bubble shell; and zone d is the ambient ISM.
    ii.~- The structure of the modified Castor et al.~bubble, showing 
    leaking from the shell. Note the larger size of region a, smaller 
    size of region b and material escaping into region d.
}} \label{shape}
\end{center}
\end{figure}

In the Castor et al.~theory the stars are assumed to blow a constant,
spherically symmetric stellar wind that interacts with the ambient ISM
and produces a bubble.  For the first few hundred years the stellar
wind freely expands at the wind velocity.  Once the wind shocks, there
is a period of adiabatic expansion for a few thousand years.  The
bubble then enters the ``snow plough'' stage when swept up mass
densities reach the critical point for radiative cooling and a thin
shell forms.  In the Castor scenario the snow plough stage is the most
readily observed phase as it is by far the longest lived.
During this time the bubble has a four zone structure,
(Fig.~\ref{shape}(i)):  {\bf a}) A hypersonic stellar wind.  {\bf b}) A
hot ($T \sim 10^6$ K), almost isobaric region consisting of shocked
stellar wind mixed with a small fraction of the swept-up interstellar
gas.  {\bf c}) A thin, dense, cold ($T << 10^6$ K) shell containing
most of the swept up material at $R_b$.  {\bf d}) Ambient
interstellar gas.

One of the most important features of the snow plough structure is the
cooling between zones b and c.  The shell, c, is dense and cold due to
efficient radiative cooling.  However, zone b is hot as its density is
too low to radiatively cool.  Thermal conduction between b and c
causes gas from the inner edge of c to evaporate into b.  Although the
mass loss from c is negligible compared to the accretion of mass from
swept up ISM, it dominates over stellar wind as a source of matter in
b.  The bubble shell is pushed outward by the pressure from the
confined hot gas.  The equations describing the evolution are given in
Appendix II.

Unlike the Chevalier \& Clegg theory, the Castor theory takes into
consideration the mass swept up from the surrounding nebula, requiring
an ISM or GMC model.  The original papers \citep{cas75,wea77,che85}
assumed a homogeneous ISM.  However, in our models for Carina we
sometimes employ an isothermal sphere as an approximation for the GMC
surrounding the cluster.  The values for the cluster gas mass and
radius, and the surrounding GMC mass and radius were taken from
\citet{mur07} assuming a density profile in the cluster of $r^{-1}$
and in the GMC of $r^{-2}$.  The `standard' ISM model used for our
bubble models was $R_{cl}=1\pc$, $M_{cl}=2.6\times10^4M_\odot$,
$R_G=24.62\pc$, and $M_G=3.2\times10^5M_\odot$,
%
%
where $M_{cl}$ is the mass in gas and dust remaining in the cluster
after star formation assuming a star formation efficiency of $\sim
35$\%.  The $R_G$ was chosen so that the density at the boundary
between the cluster and GMC was continuous.

In other cases we assume that the density falls off as $\rho(r)\sim
1/r$, or even that the density is (on average) uniform.

In the Castor models we assume that the cluster produces a point
source wind from the centre of the cluster. Although this does not
accurately represent the stellar wind within the cluster radius,
simulations by \citet{can00} show that the approximation is valid
once outside the cluster radius.

To model the Castor bubble, we used the integrate\_ode program from
Numerical Recipes \citep{numre} to simultaneously integrate the
momentum, radius, mass and energy evolution equations (equations
(\ref{cas1}) to (\ref{cas4})) from the initial conditions in
\citet{wea77} (equations (\ref{casic1}) to (\ref{casic4})).

The X-ray luminosity was calculated taking into account the radial profile
of density and temperature through the bubble \citep{wea77}.  When the
X-ray luminosity is calculated the bubble is split into 5000 shells of
equal width between $R_a$ and $R_b$ and the X-ray luminosity in the
Einstein band from Bremsstrahlung radiation is calculated for each
shell.  A resolution check was conducted at many different bubble
radii to ensure accuracy.

\subsubsection{Modifications to the Castor bubble model - stellar mass loss rates}

The generally accepted mass loss rate of O stars assumes a 
smooth (not clumpy) outflow throughout the main sequence lifetime.
However, recent observations show that stellar winds are clumpy and hence
the mass loss rates are likely lower than previously accepted
\citep{ful04,bou05,eva04,pul06}. These observations suggests that the mass loss rate
is at least three times and possibly more than 30 times lower than
previously expected.  We introduce $\alpha$ to parametrize the
reduction in the the mass loss rate: $\dot{M}_{w} = \alpha
\dot{M}_{w,original}$ and $L_w = \alpha L_{w,original}$.

\subsubsection{Radiation pressure}
The far infrared luminosity in the direction of Carina
is roughly half the total stellar luminosity. It follows that half or more
of the momentum carried by light emitted by the stars is deposited
into the gas near the stars.
Thus, the radiation pressure contributes to the outward force on the
bubble shell (where the UV optical depth is large)
\begin{equation}
F_{rad} = \eta \frac{L_{bol}}{c}  ,
\end{equation}
where $\eta$ is of order $1/2$, $c$ is the speed of light and
$L_{bol}$ is the total bolometric luminosity from the cluster.

Using a multi-wavelength analysis of the Carina Nebula \citet{smb07}
estimate dust temperatures and calculate the total mass in dust.
\citet{smb07} calculate the masses using the models of \citet{gil74}.
We employed more recent dust models \citep{lao93}; the mass in dust can be
calculated using the mass per grain, $4 \pi a^3 \rho /3$ and
luminosity per grain $= Q_e \sigma T_d^4 4 \pi a^2$
\begin{equation}
M_d = \frac{a \rho}{ 3 \sigma Q_e T_d^4} L_d ,
\end{equation}
where a = effective grain radius, $\sigma$ is the Stefan-Boltzmann
constant, $Q_e$ = mean thermal emissivity, $T_d$ = dust temperature,
and $L_d$ = the dust luminosity.  No silicate features are seen so we
can assume the grains are carbon.  We also assume that grains are
small ($a < 0.2 \mu \hbox{m}$).  The $Q_e$ values for each temperature
can be read from the graph in \citet{lao93}. In particular,
$Q_e\approx3\times10^{-3}$ for $T_d=35\K$. This is factor of $9$ larger
than the value employed by \citet{smb07}; consequently, we find a
total dust mass $M_d\approx1000(\rho/3\g\cm^{-3})M_\odot$, Table
\ref{dustmass}, a factor $9$ smaller than \citet{smb07}.

Assuming a dust to gas ratio of 1:100 the total mass for the
surrounding nebula is $\sim 10^5 M_\odot$, corresponding to an
$A_V\approx 1\, (20\pc/R_b)^2$. In contrast, \citet{smb07} find 
$9.6 \times 10^5 M_\odot$, corresponding to an $A_V\approx10\,
(20\pc/R_b)^2$. 

\citet{smb07} find $T_{D}\approx35\K$, which we argue supports the
lower mass estimate, as follows.  Using the expression for dust
temperature
\be   
T_d=49f^{-1/(4+\beta)}
\left({2\times10^{17}\cm\over r}\right)^{2/(4+\beta)}
\left({L_{bol}\over 10^5L_\odot}\right)^{1/(4+\beta)}
\ee  
from \citet{1976ApJ...206..718S} with
$f=Q_{abs}(\lambda=50\mu)\approx0.01$ we find that the typical
distance between a dust grain and Tr 16 is $\sim 24\pc$ (taking
$\beta=1$). We note that this is similar to the estimated bubble
size. Since the bulk of the stellar radiation is initially emitted in
the UV, it will be absorbed at $A_v\lesssim1$, consistent with the
lower mass estimate given above.

\subsubsection{Self-Gravity}

As the shell sweeps up a substantial amount of mass, the self-gravity
of the shell and gravity between the shell and the stars and interior
gas should be considered in the momentum equation.
\begin{equation}
F_{grav} = - \frac{G M_c^2}{2 R_b^2} -\frac{G M_c  (M_b + M_{stars})}{R_b^2} .
\end{equation}

\subsubsection{Leakage of hot gas}

The Castor theory assumes expansion through an homogeneous ISM.
However, as noted in the introduction, turbulence is expected to
produce density fluctuations in the ISM, while observations show that
the column density is log-normally distributed. Thus, sections of the
bubble shell with low column will expand faster and high column
sections will expand more slowly.  This uneven expansion will result
in gaps in the shell, allowing the hot gas to escape from the bubble.

We investigated the effects of holes in the expanding shell
using a toy model.  Our model consisted of a Castor-style
bubble with a shell cover fraction $C_f$: when
$C_f = 1$ the shell has no holes and when $C_f = 0$ the
shell is covered with holes, i.e.~no shell exists.  These holes allow the
interior gas to escape at the speed of sound, $c_s = (5 k_b T / (3
m_p))^{1/2}$,
 with a mass flux and energy flux given by
\begin{equation}
\dot{M_{b}} = - (1-C_f) 4 \pi R_b^2  \rho_{b} c_s,
\end{equation}
\begin{equation}
\dot{E_{b}} = -  (1-C_f) 4 \pi R_b^2  \frac{5}{2} \rho_{b} c_s^3,
\end{equation}
where $R_b$ is the radius of the bubble interior, $\rho_b$ is the density
inside the shell, which is assumed to be homogeneous and $\gamma = 5/3$ throught this paper.


We are not concerned with the bubble evolution within the cluster
radius (of order $3\pc$ for Tr 16).  Thus, we neglect gravity and
radiation pressure for  $r<R_{cl}$; once the bubble expands, under the
influence of the other forces, beyond this radius, we smoothly turn on
gravity and radiation pressure.

With hot gas from the interior able to leak away, it is important to
consider the location of the stellar wind shock, $R_a$ (see
Fig.~\ref{shape}): if $R_a$ approaches the shell radius at any time
our model breaks down.  We use the assumed ambient ISM density profile
for the regions where the shell is still intact and negligible density
where there are holes.  Although this is not an accurate description
of a clumpy ISM, it does allow the shell to leak while still gaining
mass as it expands through a variable density ISM.  The gas and
radiation pressure only acts on the remaining shell (i.e.~area = $C_f
4 \pi R_b^2$ ). 

Without leaking the radiative losses in region b were negligible. However,
with leaking this may not always be the case. To ensure these losses are 
taken into account when significant we calculate the radiative luminosity 
of region b, $L_b$, and include it in the energy equation.

The sizes of the holes in the shell may vary as the bubble expands, so
that $C_f=C_f(t)$ or $C_f(r)$.
To investigate the effects of different hole expansion models, we
conducted runs with many different power law relationships with time
(including  $C_f(t)=const.$), which we discuss below.

\subsubsection{Final bubble model}

Considering radiation pressure, gas pressure, stellar winds, gravity
and leaking the evolution equations for momentum, radius, mass and energy become
\begin{eqnarray}
\frac{d{\cal P}_c}{dt} & = & 4 \pi R_b^2 P_b + \frac{L_{bol}}{c} - 
\frac{G M_c^2}{2 R_b^2} -\frac{G M_c (M_b+ M_*)}{ R_b^2} ,\label{mom_eqn} \\
\frac{dR_b}{dt} & = & \frac{{\cal P}_c}{M_c}  ,\label{rad_eqn} \\
\frac{dM_b}{dt} & = & C_1 T_b^{5/2} R_b^2 (R_b - R_a)^{-1} - 10^{-5} 
\frac{\mu}{k_b} L_b - (1-C_f) 4 \pi R_b^2 \rho_b c_s,\label{mass_eqn} \\
\frac{dE_b}{dt} & = & L_w - 4 \pi R_b^2 P_b \frac{dR_b}{dt} - L_b - \frac{5}{2} (1-C_f) 4 \pi R_b^2 \rho_b c_s^3 .\label{energy_eqn} 
\end{eqnarray}
respectively, see Fig.~\ref{shape}(ii).

${\cal P}_c$ is the momentum of the shell,
$t$ is time,
$R_b$ is the radius of the bubble,
$P_b$ is the gas pressure in the bubble interior,
$L_{bol}$ is the total bolometric luminosity of the stars, 
$c$ is the speed of light,
$G$ is the gravitational constant,
$M_c$ is the mass of gas in the bubble shell,
$M_b$ is the mass of gas in the bubble interior,
$M_*=10^4 M_{sun}$ is the mass of stars in the cluster,
$C_1$ is a combination of constants associated with conduction,
$T_b$ is the temperature of the gas in the bubble interior,
$R_a$ is the radius of the stellar wind shock,
$k_b$ is the Boltzmann constant,
$L_b$ is the luminosity emitted from the bubble interior,
$C_f$ is the covering fraction of the shell,
$\rho_b$ is the mass density of the bubble interior,
$c_s$ is the speed of sound of the bubble interior,
$E_b$ is the energy in the gas in the bubble interior,
$L_w$ is the stellar wind luminosity.

The terms on the right hand side of the momentum equation
(\ref{mom_eqn}) represent, in order, the outward force from the gas
pressure, the outward force from radiation, the inward force from the
self gravity of the bubble shell, and the inward force from gravity
between the star cluster (with a minor contribution from the gas in
the bubble interior) and the bubble shell.  The terms on the right
hand side of the mass equation (\ref{mass_eqn}) are the mass input
associated with conduction evaporating the bubble wall, a (downward)
correction in the conductive mass loss rate associated with radiative
cooling, and mass loss from hot gas escaping the bubble.  The terms on
the right hand side of the energy equation (\ref{energy_eqn}) are
stellar wind luminosity, adiabatic expansion, radiative cooling, and
energy advection associated with the loss of hot gas through the
bubble wall.

\section{RESULTS \label{sec:results}}

\subsection{Original stellar wind models \label{sec:original}}

\begin{figure}
\epsscale{.90}
\plotone{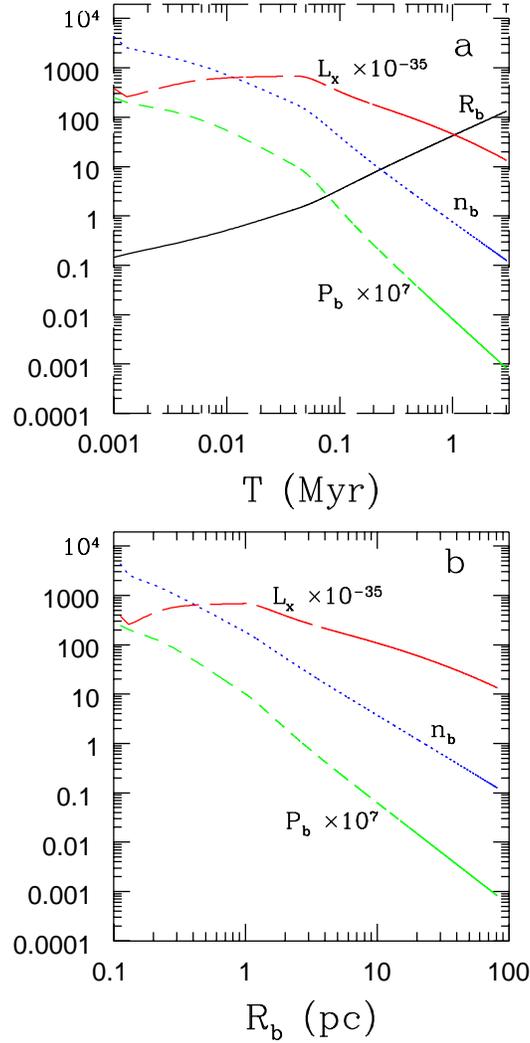} 
\caption{{\small Plots of the basic \citet{cas75} model evolution
    against time (a) and $R_b$ (b) for Tr 16. These plots show how
    pressure, number density and total X-ray luminosity change through
    the evolution of the bubble. Pressure and X-ray luminosity have been scaled to
    fit on the graphs.} }
\label{casplot}
\end{figure}

\begin{figure}
\begin{center}
\epsscale{0.9}
\plotone{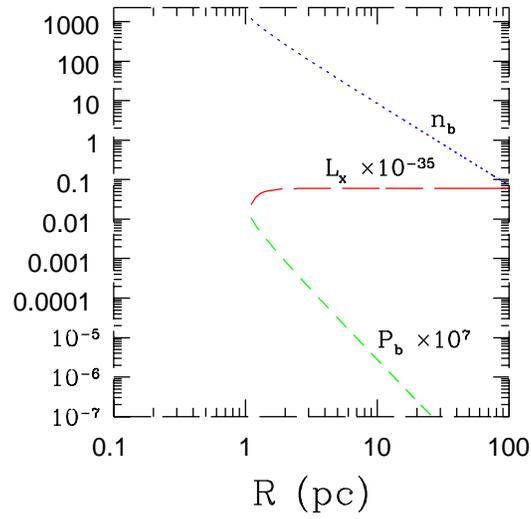} 
\end{center}
\caption{{\small Plot of the steady state \citet{che85} model for Tr
    16. The number density, pressure and X-ray luminosity against
    radius are plotted. Pressure and X-ray luminosity have been scaled to
    fit on the graphs. Note that we plot the {\em cumulative} X-ray luminosity
    as a function of radius.} }
\label{chevplot}
\end{figure}

The Chevalier \& Clegg model predicts a free-free X-ray luminosity in
the Einstein band ($0.5 - 3 \keV$) of $ L_x = 3.75 \times
10^{33}\ergs$ for Carina. The Castor model predicts an X-ray
luminosity of $L_x = 5.4 \times 10^{36}\ergs$ for $R_b=13\pc$ and
$L_x = 3.8 \times 10^{36}\ergs$ for $R_b=20\pc$.  Plots of the models
can be seen in Fig.~\ref{casplot} and Fig.~\ref{chevplot}.  Both of
these models considered X-ray emission from $r > R_{cl}$ to match the
observations, in which the cluster region is ignored due to
contaminating stellar and wind-wind collisional X-ray emission. 

In the Chevalier \& Clegg model the inner region of the wind dominates
the X-ray luminosity so a strong radial gradient in the X-ray surface
brightness is predicted; beyond a few parsecs the X-ray luminosity is
negligible. In contrast, in the Castor et al.~model the density is almost
homogeneous within the bubble, so very little radial surface
brightness gradient is expected, aside from a sharp edge at $R_b$.

\begin{figure}
\begin{center}
\epsscale{0.9} 
\plotone{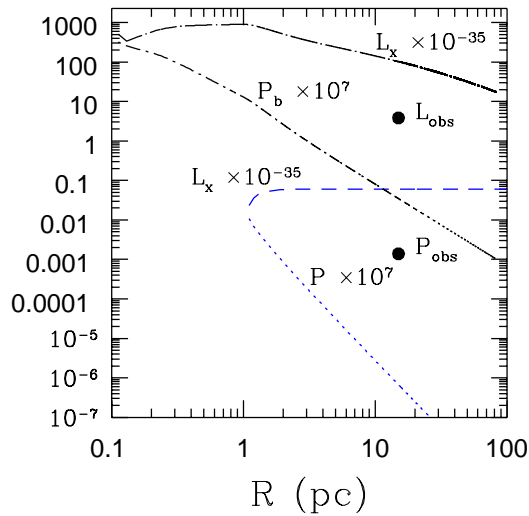}
\end{center}
\caption{{\small Plot of the scaled X-ray luminosity and pressure calculated
    as a function of radius for the two classes of (unmodified) models
    for Tr 16, as discussed in \S \ref{sec:original}. The Castor model is 
    plotted as a function of bubble size whereas the Chevalier model is 
    plotted as a cumulative function of radius (due to its steady state nature). 
    The upper L$_x$ and P lines are for the Castor model and the lower lines for the
    Chevalier model. The two dots show the observed pressure (lower) and X-ray
    luminosity (upper). In the Castor et al.~model the pressure is roughly
    constant for $r<R_b$, but this constant pressure is a function of
    the bubble radius, as shown. In the Chevalier \& Clegg model there
    is no bubble, but pressure decreases with increasing radius.} }
\label{chevcas}
\end{figure}

Observations of diffuse X-ray emission in Carina give $L_x = 6.1
\times 10^{34}\ergs$ \citep{sew79}; those authors assumed the distance
to the cluster was $2.6\kpc$.  If we assume the distance is $2.3\kpc$,
the luminosity becomes $4.8\times10^{34}\ergs$.  The observed X-ray
emission is ten times higher than predicted by Chevalier \& Clegg
model but one hundred times lower than predicted by the Castor model,
Fig.~\ref{chevcas}.  The observed pressure, as estimated by `the
finger' and radio emission, is closer to the calculated Castor
pressure when the bubble has the currently observed radius
$R_b\approx20\pc$.

As noted elsewhere, the Chevalier \& Clegg model predicts a pressure
that has a strong radial gradient, so the pressure is dependent upon
the location within the bubble rather than the bubble size. However,
the Chevalier \& Clegg pressure is far below the observed pressure at
most points in the bubble.

It is unclear how long it takes the bubble to reach the cluster
radius, as the expansion time will depend upon the rather complicated
wind-wind interactions within the cluster.  To properly understand the
early evolution of bubbles, full hydrodynamical models are needed.
However, for now we will assume it is much less than a million years,
as material will be displaced during star formation due to pre-main
sequence jets and outflows.  We assume that the majority of the time
taken to expand to the observed size is accumulated when $R_{cl} < r <
R_{obs}$.  Using this assumption we found that to expand to 20 parsecs
the Castor model took only $0.90$ million years, far shorter than the
estimated age of Carina (around $3.6$ Myrs).

It follows that the situation for the Castor et al.~model is actually
worse than the impression conveyed by the Figure, for the following
reason. The actual age of the bubble is a factor of four higher than
in the model, so the energy content of the bubble is underestimated in
the model by the same factor, as is the pressure. Given the weak
dependence of the temperature on time, the X-ray luminosity scales as
the square of the pressure; if the radius, wind luminosity, and age of
the bubble are set by observations, then the X-ray luminosity would be
a factor of $16$ larger than in the model.

\subsection{Modified Castor et al.~model; leakage \label{sec:leaky}}

Taken together, the overestimate of the pressure and X-ray flux, and
the underestimate of the bubble age (or overestimate of the bubble
size at a fixed age) suggest that the outward force in the Castor et
al. model is overestimated.

The inclusion of radiation pressure and gravity for the Tr 16 model
makes only a small difference to the evolution of the bubble.  For Tr
16 $ |F_{HII}| \approx |F_{RP}| \approx |F_{grav}|$; in the model,
unlike the case in Carina, all are much smaller than the gas-pressure
force exerted by the shocked stellar wind, assuming no leakage and a
standard O star mass-loss rate.

In an attempt to produce models in closer agreement with observations,
it is clear that the internal pressure of the bubble must be
reduced. One way to do this is to allow hot gas to leak out of the
bubble ($C_f<1$), or equivalently to allow energy to leak out of
the hot gas, via radiation or conduction; another is to reduce the
mass loss rate and hence the kinetic luminosity of the stellar winds
($\alpha<1$).

Observation of massive clusters support the possibility of leaking. 
Firstly, for Carina and many other clusters you can see the stars, 
showing that there must be a gap in the shell along our line of sight
\citep{hubca}. Secondly, recent observations of M17 and RCW49 by 
\citet{pov08} show stellar wind bow shocks around O stars at the edge 
and outside of the bubbles suggestive of large scale gas outflow from the 
H II region (see their Figure 2, especially RCW49-S1).

\begin{figure}
\epsscale{.9}
\plotone{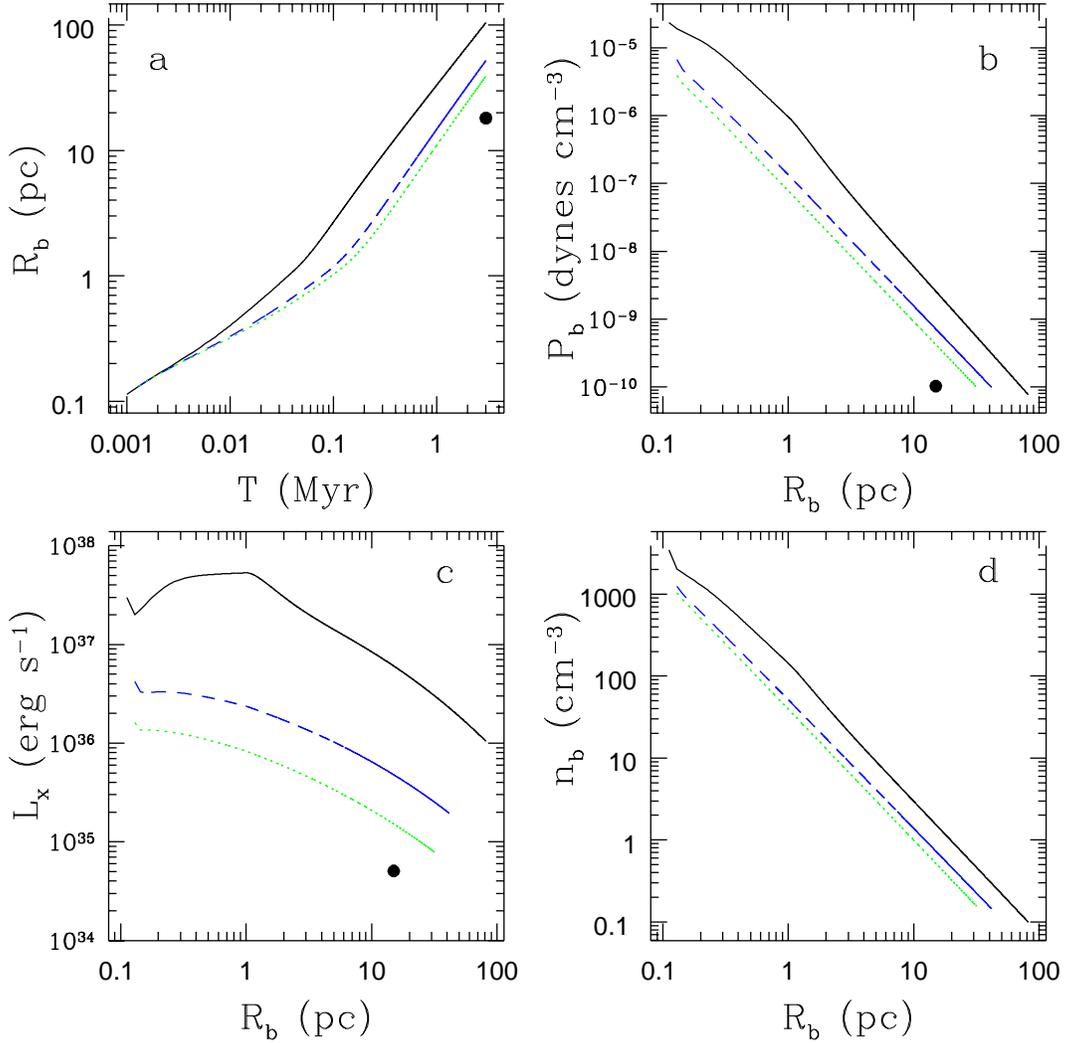}
\caption{ Graphs for three modified Castor simulations with $C_f
  = 1$ (solid), $C_f = 0.65$ (dashed), and $C_f = 0.3$
  (dotted). (a) radius versus time, (b) pressure versus
  radius, (c) X-ray luminosity versus radius, and (d) number density
  inside the bubble versus radius.  Observed results are shown as a black dot. See the
  discussion in \S \ref{sec:leaky}. \label{Fig:leaky} }
\end{figure}

Our initial simulations of a leaking bubble looked at three examples
in detail: $C_f = 1.0$, $C_f = 0.65$, and $C_f = 0.3$.
The results are given in Fig.~\ref{Fig:leaky}.
Not surprisingly, these simulations reveal that as
$C_f$ decreases the shell expands more slowly, the bubble
interior pressure is lower, as is the predicted X-ray luminosity.

The effects of varying $C_f$ with time were then investigated.
These simulations started with $C_f=0.6$ at $t = 10^3$ years and
evolved $C_f(t)$ with different power laws until $t = 3$Myrs,
when $C_f=0.2$.  The simulations show that expansion rate,
pressure, temperature and density differ very little with different
power-law exponents; however, the X-ray luminosity does depend on the
exponent. The difference in X-ray luminosity is due to the
combination of the small changes in the parameters giving larger X-ray
luminosities for larger exponents; larger exponents imply
$C_f$ being large for longer times.  Thus, less material escapes over
the lifetime of the bubble: $\dot{M}_{escape} \propto (1-C_f)$ and
the X-ray luminosity is higher during the expansion when the power-law
exponent is larger.

\begin{figure}
\plotone{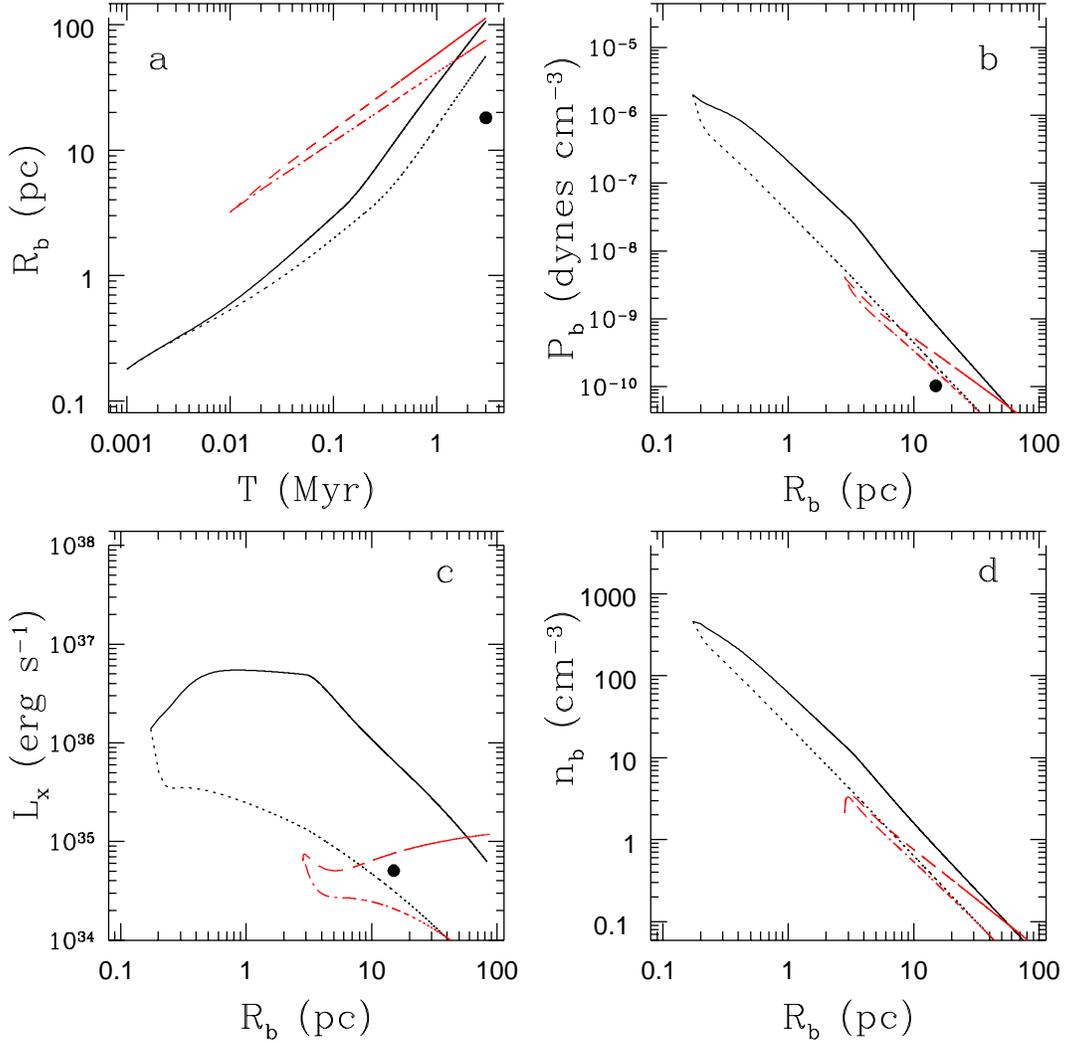}
\caption{ Graphs showing the details of modified Castor simulations
  with different ISM models and different shell cover factors $C_f$.  All runs have
  $R_{cl} = 3 \pc$, $R_G = 73.86 \pc$ and $\alpha = 1/3$. (a) is $R_{b}$
  versus  time, (b) pressure versus $R_b$, (c) X-ray luminosity versus
  $R_b$, and (d) $n_b$ versus $R_b$. The solid line
  is for an isothermal GMC, with $C_f = 1$, the dotted line is
  an isothermal GMC with $C_f = 0.5$, the dashed line has constant ISM
  density with $C_f = 1$, and the dash-dotted line has constant
  ISM density with $C_f = 0.5$.  Note that (b) and (c) are plotted
  against radius so although some models do fit the observed
  quantities at that radius they do so after too short a time.
   \label{Fig:eightruns} }
\end{figure}

To see the way $C_f$ and the run of density in the ISM affected
the evolution of the bubble we ran four different models.  We
considered $C_f=1.0$ or $0.5$ and either isothermal or constant
density models (with the same total mass), Fig.~\ref{Fig:eightruns}.  From
this study we find that a homogeneous medium leads to a faster
expansion; the mass swept up in the bubble shell is less at a given
radius in the constant density model.  However, the isothermal sphere
model does approach the homogeneous model near the end of the
simulations when both reach $R_G$ and hence the same shell
mass.  The models with $C_f=0.5$ expanded more slowly and had
significantly lower X-ray luminosities than models with $C_f=1.0$,
as expected.

\begin{figure}
\plotone{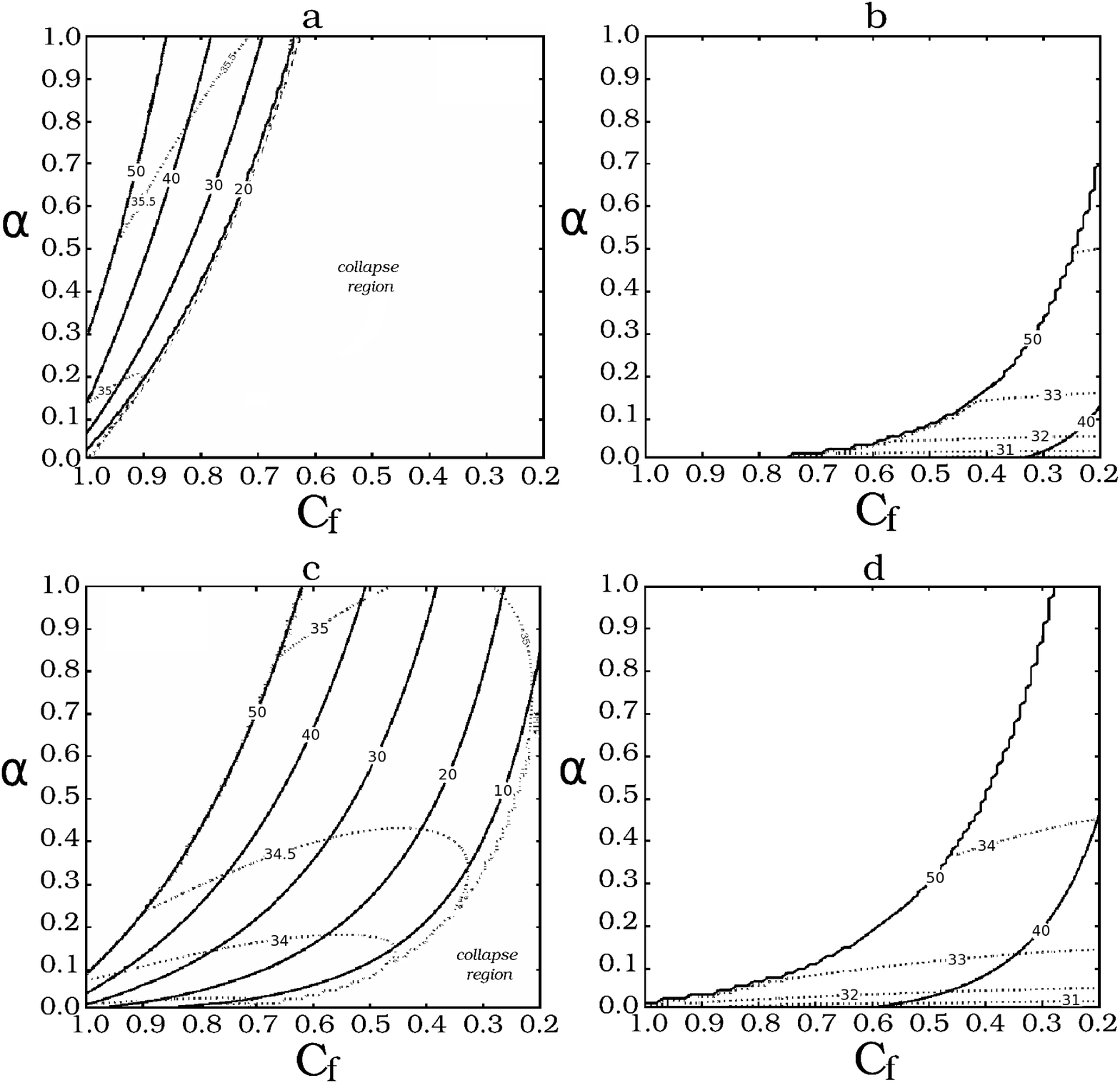}
\caption{ Parameter study results for varying $\alpha$ and $C_f$
  for either an isothermal sphere ISM and for $R_G = 24 \pc$ (a)
  or $R_G = 60 \pc$ (c), or for a constant density ISM for $R_G=24$ (b)
  or $R_G=60\pc$ (d) modified Castor models at
  3Myrs. The contours show the X-ray luminosity (dotted) and the
  radius (solid). Gray dashed lines show the boundary of the
  collapse region. X-ray luminosity is in logarithmic contours in
  intervals of $\times$10 erg s$^{-1}$ for the homogeneous models and
  $\times$10$^{0.5}$ erg s$^{-1}$ for the isothermal models. Radius contours
  are linear in intervals of 10 parsecs. The observed properties of
  Carina are $13 < R_b < 20 \pc$ and $L_x = 4.8 \times 10^{34} \ergs$. For
  the isothermal ISM model there are regions of the parameter space
  where the shell collapses inwards in the simulations. This is due to
  the swept up mass being so large the gravitational force is larger
  than the sum of the gas pressure force and radiation pressure
  force.\label{parameterone} }

\end{figure}

We conducted a parameter study in $\alpha$ and $C_f$ to see if
any combination would result in acceptable fits to the observed bubble
in Carina.  Models were calculated for $ 0.15 \leq C_f \leq 0.99$
and $0.01 \leq \alpha \leq 1.00$, stepping both parameters at
intervals of $0.01$.  The results are shown in
Fig.~\ref{parameterone};
contours are only plotted for $R_b < 50 \pc$ at a bubble age of $3$
Myrs.  This parameter space study was run for four different ISM
models.  
Runs shown in part (a) of the Figure are for an
isothermal sphere model with $R_G = 24\pc$; run (b) shows an
constant density model with $R_G = 24\pc$; run (c) is for an isothermal
sphere model with $R_G = 60\pc$; and run (d) is for a constant density
model with $R_G = 60\pc$.  

The results show that isothermal sphere models gives smaller radii and
slightly higher X-ray luminosity for any given $(\alpha,C_f)$
pair.  The final radius at $3$ million years increases with increasing
$\alpha$ and with increasing $C_f$, as expected.  For the
isothermal ISM model the shell actually collapses inwards for low
values of $\alpha$ and low values of $C_f$, a result of the self
gravity of the shell exerting an inward force larger than the sum of
the gas pressure force and radiation pressure force on the shell.

The X-ray luminosity increases with increasing $\alpha$ and increasing
$C_f$, again as expected.  The region of parameter space that
best fits the observations is for an isothermal sphere model with $R_G
= 60\pc$.  For this class of model there is a large parameter space
with $R_b$ between $10$ and $20$ parsecs
and X-ray luminosities less than $4.8\times 10^{34}\ergs$.  For example, for
$\alpha = 0.2$ and $C_f = 0.5$, $R_b = 16.5\pc$, $L_x = 1.2 \times
10^{34}\ergs$, and $P_b = 1.1 \times 10^{-10}\dynes\,\cm^{-2}$.

\section{X-RAY SURFACE BRIGHTNESS PROFILES}
\label{sec:globules}

\begin{figure}
\begin{center}
\epsscale{.90}
\plotone{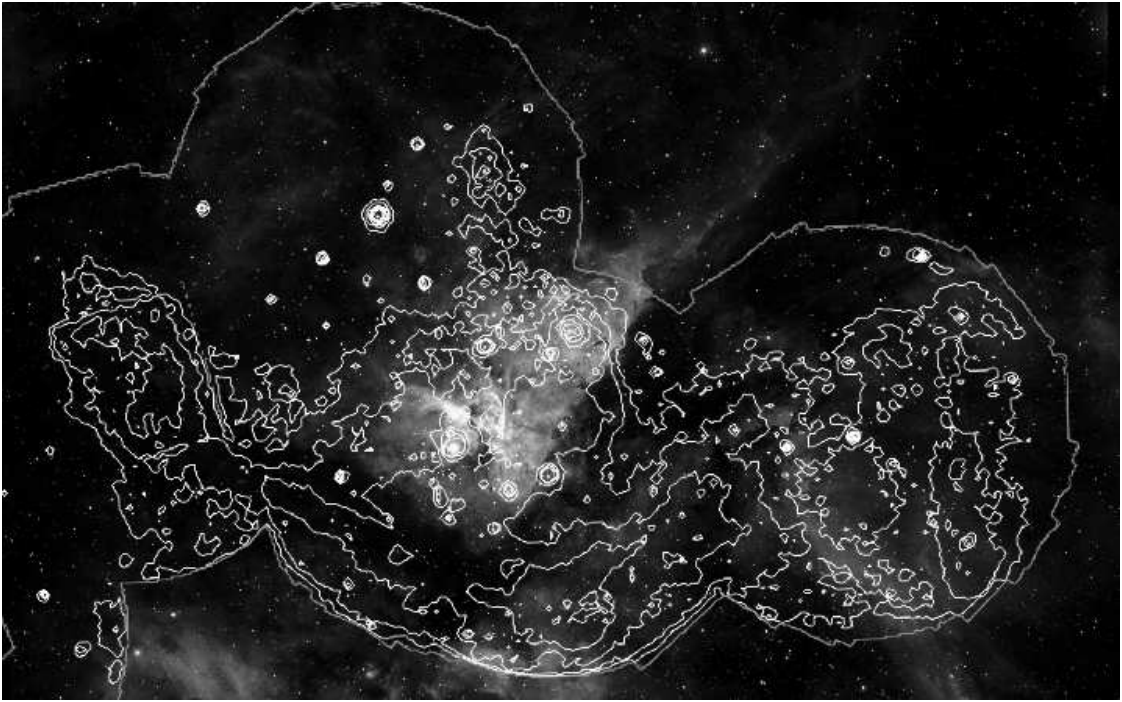} 
\end{center}
\caption{{\small Image of the Carina nebula in visual \citep{smb07} and X-ray
    contours \citep{ham07}. The visual image is a combination of H II, O IV and S
    II and the X-ray image is between $0.7\keV$ and $1.3\keV$. The green
    outline shows the edge of the X-ray image. See the electronic
    edition of the Journal for a color version of this Figure.} }
\label{hubxmm}
\end{figure}

The morphology of the X-ray emission provides a diagnostic to test our
models. Matching up the X-ray emission \citep{ham07} and the visual
images of the Carina nebula \citep{hubca}, Fig.~\ref{hubxmm}, it can
be seen that the X-ray surface brightness increases near the edge of the
bubble, the latter seen as a silhouette in the visible image.  This
strongly suggests that a significant fraction of the diffuse X-ray
emission seen by \citet{sew79} was produced by gas evaporating from the
bubble wall.

\begin{figure}
\plotone{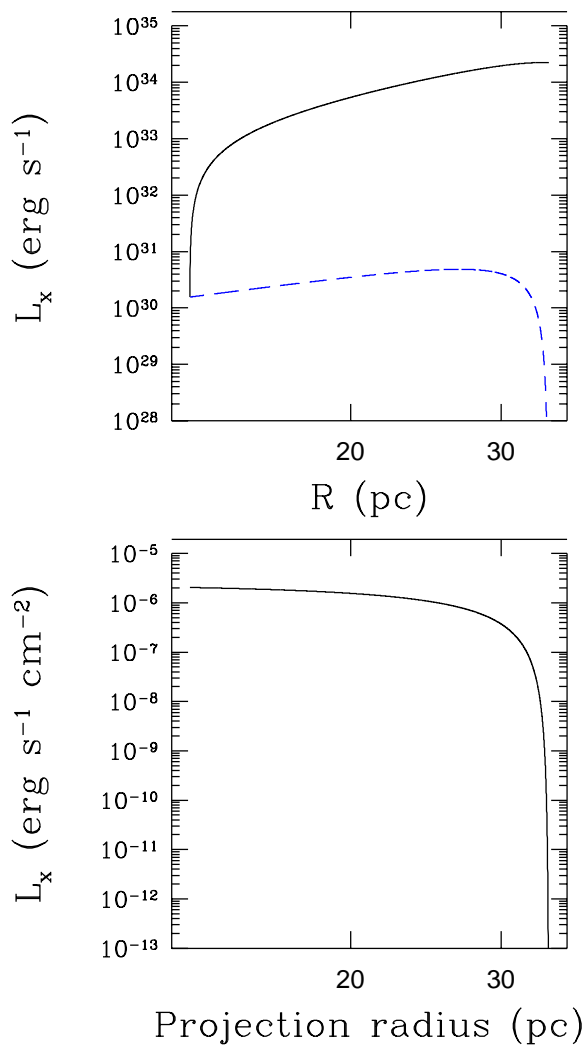}
\caption{ Graphs of a modified Castor model showing (a) the
    cumulative X-ray luminosity (solid) and X-ray profile
    (dashed) with 3D radius; (b) shows the X-ray profile with
    projected radius (z);  we expect an almost
    flat X-ray profile with a sharp cutoff at the bubble edge. This
    model is for $\alpha = 1/3$, $C_f = 0.5$, isothermal sphere
    ISM model with $R_G = 60 \pc$. }
\label{xraystruc}
\end{figure}

This surface brightness profile is not predicted by either the
\citet{che85} or \citet{cas75} models; the former predicts that
surface brightness falls monotonically with increasing $r$, while the
latter predicts a nearly flat surface brightness profile out to
$r=R_b$.  For an example of the latter, consider one run with $\alpha
= 1/3$, $C_f = 0.5$, an isothermal sphere ISM with $R_G = 60\pc$,
and stellar wind, radiation pressure, and gravity all included. At an
age of $3$ Myrs we find a bubble radius of $34\pc$ and a total X-ray
luminosity of $1.3 \times 10^{34}\ergs$, roughly consistent with the
observed bubble.  The cumulative X-ray emission as a function of the 
3D (not projected) radius is shown in Fig.~\ref{xraystruc}(a); the X-ray
profile with 2D (projected) radius is shown in Fig.~\ref{xraystruc}(b).
The calculated profiles show an almost flat X-ray profile with a
sharp edge at the bubble boundary.  This contrasts with the \citet{sew79}
and more recent XMM images \citep{ham07}, which are
much more complex, with an elevated surface brightness near the bubble
walls as noted above.

\subsection{Globule evaporation}

To estimate the emission from the interface between the hot stellar
wind material and the molecular gas around the bubble, we first look
at a simplified case of a spherical globule of cold gas embedded
within the hot gas of the bubble interior, the latter having
temperature $T_f$ and number density $n_f$ at large distances from the
cold globule. The cold gas will be heated by conduction and evaporate,
as discussed in, e.g., \citet{cow77}.  
The theory makes three assumptions; firstly the globule is large
enough to give a roughly time-independent solution for the mass loss
rate, secondly that the thermal conduction is unsaturated, and  thirdly that the flow has a
moderate Mach number ($\mathcal{M}^2 << 5$).
The conduction is unsaturated if the thermal mean free path for electrons is much shorter than the temperature scale height.
For the Carina
nebula, with $n_x\approx0.1\cm^{-3}$ and $T_x=6\times10^6\K$,
the conductivity is approaching the saturated limit for globule radius
$R_1 \approx 1\pc$.
Assuming we are in the unsaturated case, the mass loss rate is given by the classical equation (same as equation 2)
\begin{equation}
\dot{m} = \frac{16 \pi \mu K(T) R_1}{25k_b}
\end{equation}
where $\mu$ is the mean mass per particle, $K(T)$ is the conductivity
at $T_f$, and $R_1$ is the radius of the globule.   The dimensionless form of the equation of motion is
\begin{equation}
 (1-\mathcal{M}^2) \frac{d \ln \mathcal{M}^2}{dy} =  
\frac{2(6-5y) + \mathcal{M}^2 -1}{2.5 y(y-1)},
\end{equation}
where $\mathcal{M}$ is the mach number of the flow and $y = r/R_1$.
We assume that the flow will shock somewhere close
to the sonic point and thus still satisfy $\mathcal{M}^2 << 5$ at all
radii.

The solution of the energy equation gives the
temperature profile
\begin{equation}
T(y) = T_f \left(1- y^{-1}\right)^{2/5} ,
\end{equation}
assuming the globule temperature $T<< T_f$.  The local sound speed can
be calculated from the temperature profile and thus the velocity
deduced from the local Mach number.  The local density is then
\begin{equation} \label{eqn:density}
n(y) = \frac{\dot{m}}{4 \pi y^2 R_1^2 c_s \mathcal{M}} .
\end{equation}

\begin{figure}
\begin{center}
\epsscale{.80}
\plotone{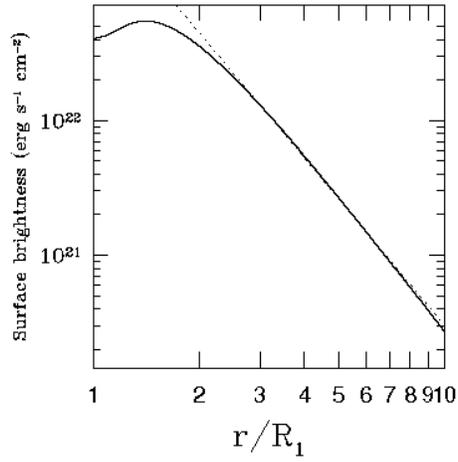} 
\end{center}
\caption{{\small The X-ray emission profile against $r/R_1$ 
    for $T_f = 10^6$ and $R_1 = 1\pc$ for an evaporating
    globule (solid line) showing a fairly sharply defined X-ray halo. 
    The dotted line shows a gradient of -3. } }
\label{xrayout}
\end{figure}

The run of temperature and density can be combined with the expression
for Bremsstrahlung X-ray emission to give the expected X-ray
luminosity and surface brightness for an evaporating globule.  
The intensity along a line of sight with an impact parameter
$b$ to a globule is proportional to 
\be   
\int dz n^2(z) T^{1/2}(z)e^{-kT/h\nu}\approx 
\int dz {T^{1/2}(r(z))\over (b^2+z^2)^2}e^{-kT/h\nu}\sim \left( \frac{R_1}{b} \right)^3
\ee  
where $z$ is measured along the line of sight, $z=0$ in the plane of
the sky at the location of the globule, and $r=\sqrt{b^2+z^2}$. Once
$T(r)$ approaches $h\nu/k$, the surface brightness depends primarily
on $b$, falling rapidly with increasing $b$ as long as $R_1<b$.  Doing
the integral numerically, setting $T_f = 6\times 10^6$ and $R_1 =
1\pc$ we get $\dot{m} = 6 \times 10^{21}$g s$^{-1}$ and the X-ray emission
profile shown in Fig.~\ref{xrayout}; the dotted line shows the
$b^{-3}$ scaling.  In words, a spherical globule should be surrounded
by a fairly sharply defined X-ray halo.

As a check, the dependence of the total X-ray luminosity and mass loss
rate on $T_f$ and $R_1$ were investigated.  We varied the external
temperature for a globule of $R_1 = 1\pc$; in a separate set of
integrations at fixed $T_f = 10^6$K we varied the radius of the
globule $R_1$.   These simulations show that $\dot{M} \propto
T^{2.5}$, $L_x \propto T$, and $\dot{M} \propto T$ as expected
analytically for an unsaturated conductivity.

A very crude calculation of the X-ray emission produced by the
conductive evaporation of a collection of
globules forming the bubble wall suggests that most or all of the
X-ray emission could arise in this manner: 
\be  
L_x=\int 4\pi r^2 \Lambda_x n^2(r)dr.
\ee 
Using eqn. (\ref{eqn:density}),
\be 
L_x\approx4\pi\Lambda_x\left({\dot m\over 4\pi c_s m_p}\right)^2
{1\over R_1}\approx2.6\times10^{33}\left({R_1\over 1\pc}\right)
\left({\Lambda_x\over 3\times10^{-23}\ergs\cm^3}\right)\ergs.
\ee  
If the bubble wall (or about half of it) were tiled with $R_1\sim3\pc$
globules, evaporation from the wall could explain the observed X-ray
emission.

In Carina the situation is much more complicated than this toy
model. The density of the hot gas is manifestly not uniform, and in
addition we have seen that hot gas apparently escapes rapidly from the
bubble interior. We are assuming that this implies that the gas density
decreases rapidly away from the bubble wall, on a scale comparable to
the size of the wall fragments, which are typically much smaller than
the radius of the bubble. Further work along these lines is clearly
needed.

\begin{figure}
\begin{center}
\epsscale{.80}
\plotone{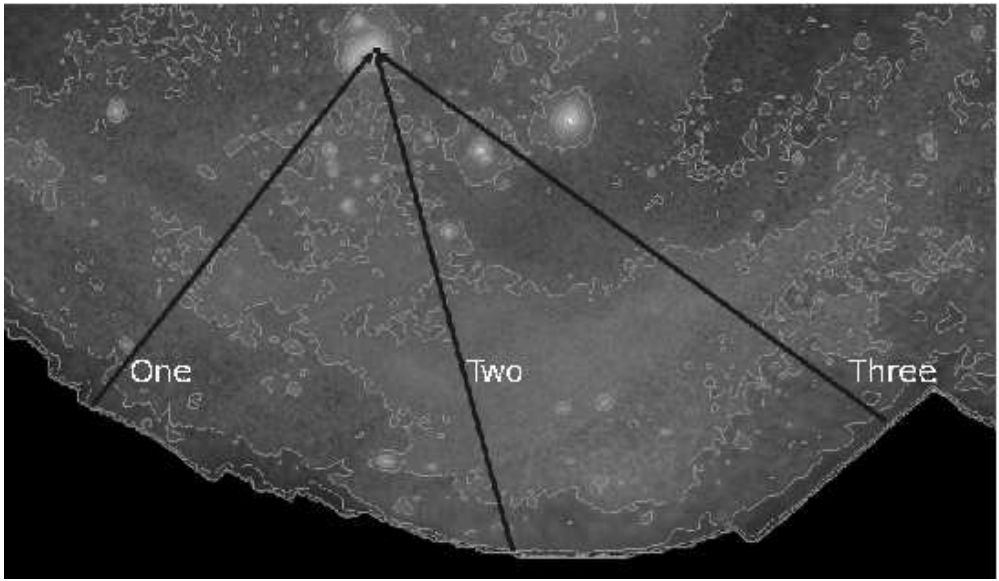} 
\end{center}
\caption{{\small XMM image of the southern part of the Carina Nebula
    between 700 eV and 1300 eV with contours. $\eta\ $ Carina can be seen
    along the top (with all three profile lines pointing at it). Thick
    lines are the profile lines used in Fig.~\ref{sblines}. }}
\label{sbpic}
\end{figure}

\begin{figure}
\begin{center}
\begin{tabular}{cc}
\plotone{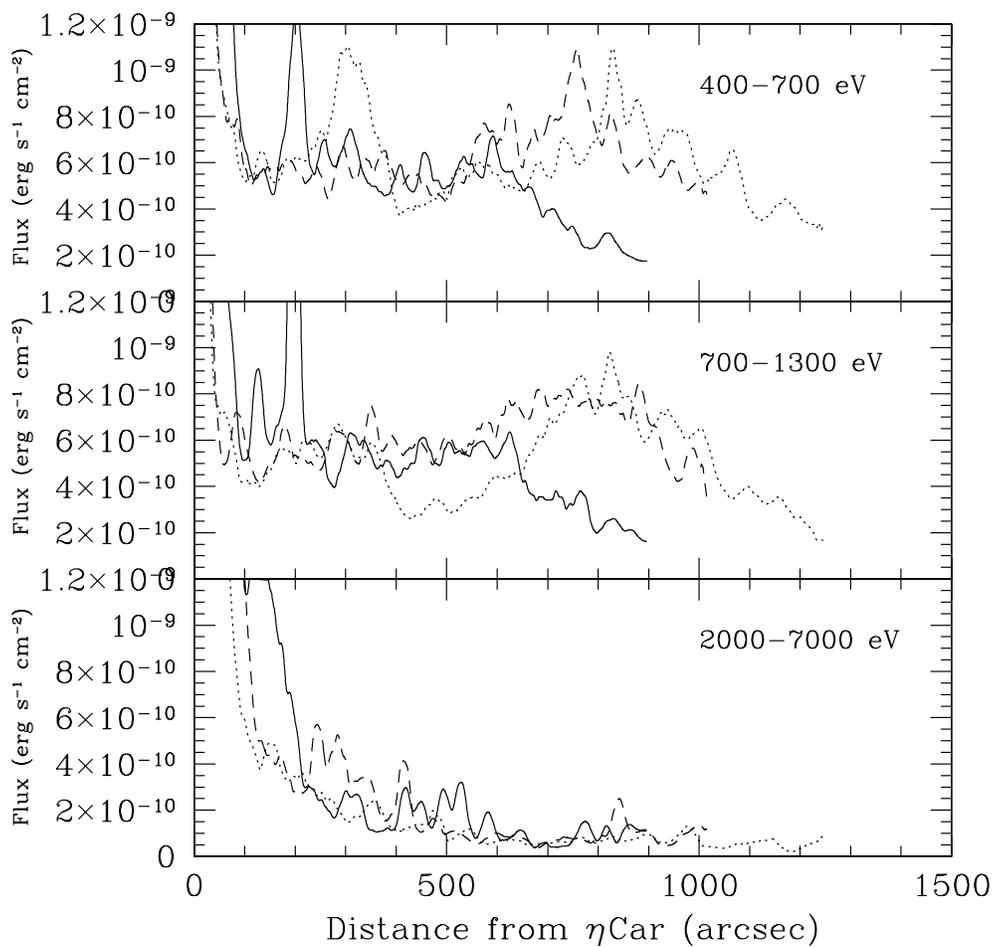}
\end{tabular}
\end{center}
\caption{{\small Surface brightness along the lines in
    Fig.~\ref{sbpic} for the three different XMM bands. 
    The solid line is for profile one, the dashed line 
    for profile two and dotted line for profile three. Surface
    brightness is given in $\erg\cm^{-2}\s^{-1}$, with an assumed $N_H =
    1.8 \times 10^{21}\cm^{-2}$ \citep{ham07}; projected distances from $\eta\ $
    Carina are given in arcseconds. Recall that at a distance of
    $2.3\kpc$, $1000$ arcseconds is about $11\pc$.}}
\label{sblines}
\end{figure}

We have extracted surface brightness profiles along rays emanating
from $\eta\ $ Carina using the XMM observations of \citep{ham07}.  The
southern part of the XMM image is shown in Fig.~\ref{sbpic}, which also
shows the three rays along which the surface brightness was
calculated for Fig.~\ref{sblines}. These profiles were produced by interpolating through the
pixels using IDL and then smoothing.  All three lines were
interpolated through 2554 points and then smoothed by a box car average
with a width of 60 points.

Emission was converted from counts to flux using the PIMMS package.
We assumed a spectrum of two black bodies as found by \citet{ham07}.

The 700-1300eV band shows the greatest contrast, from little emission
in the bubble interior (but away from $\eta\ $ Carina) to substantial
emission near the optically dark lanes. The 400-700eV band also shows
some edge effects, especially along lines Two and Three.  The high
energy band, $2000-7000{\rm\, eV}$, shows essentially no enhanced
emission near the bubble walls.

In the 700-1300eV band line One shows a decrease in surface brightness
from $\eta\ $ Carina out to 200 arcsec, a fairly flat profile between 200
and 650 arcsec, followed by a decrease out to the edge of the image.
Line Two shows a similar flat profile between 150 and 400 arcsec,
decreases to a minimum surface brightness at 500 arcsec, then
increases to a peak at about 900 arcsec.  The profile along line Three
is similar to line two but with a less pronounced minimum, and a peak
closer to $\eta\ $ Carina (at 750 arcsec).  The peak along ray Three is
wider than that along ray Two as it makes a larger angle with the
bubble edge normal.  These profiles quantify the visual impression
that the bulk of the bubble interior has a fairly flat X-ray surface
brightness but that there is strong emission from the edges.  The
edge profiles are of significantly different shape to that expected
from a Castor bubble, Fig.~\ref{xraystruc}(a) and are more
curved than would be expected from globule evaporation,
Fig.~\ref{xrayout}.

\section{DISCUSSION \label{sec:discussion}}
We have noted three discrepancies between the Castor et al.~scenario
and observed bubbles: the predicted bubble radii are too large at a
given cluster or stellar age, the predicted X-ray luminosities are too
high by a factor of $100$ or more, and concomitantly, the predicted
pressure of the hot gas is too large by a factor of $10$ or
more. Finally, we have  noted that the radio free-free  emission is
consistent with a filling factor of H II gas that is of order unity,
suggesting that the pressure in the bubbles is controlled by the H II
gas rather than that of the shocked stellar wind gas.

A number of authors have noted the small observed $R_{b}$, roughly by
a factor of $5$, compared to predictions
\citep{Dorland86,DM87,Oey96,Rauw02,Dunne03,SSN05}. These authors have
also noted the deficit of X-ray luminosity or hot gas pressure
compared to predictions. Suggested resolutions involve one or more of
the following: lower stellar wind luminosities, mass loss from the
bubble, or energy loss from the bubble, e.g., from super-Spitzer
effective conductivities, or highly efficient mass loading so that the
shocked stellar wind cools by conduction below X-ray temperatures.

\citet{cas75} and \citet{wea77} make three assumptions; first, that the
energy deposited by the stellar winds (or $5/11$ of it) is stored in
the hot gas inside the bubble, second, that the hot gas mass is
controlled by conduction (at the Spitzer rate), and third, that the
surrounding ISM is uniform (represented by their constant density
$n_0$). Applying the model requires knowledge of $L_w$, $n_0$, and the
age of the system. The predicted bubble radii depend only weakly on
$L_w$ and $n_0$, while the ages of star clusters are reasonably well
constrained (to within a factor of $2-4$). Since the bubble radii are
universally overestimated, at least one assumption must be incorrect.

We can acknowledge the fact that the dynamics of the model are wrong,
but still test the first assumption, that the energy deposited by the
stellar winds is stored inside the bubble. We will also assume that
conduction drives gas from the bubble wall into the interior until
$T_b\sim10^6-10^7\K$; this assumption is bolstered by the observed gas
temperatures, which cluster around $6\times10^6\K$.

The argument is that the time integrated wind luminosity deposits an
energy $E=(5/11)L_w\tau$ in the bubble interior, and that the
corresponding pressure is $2E/3V$, where the volume is assumed to be a
sphere with the observed bubble radius. Together with the assumption
that the temperature is $T\approx5\times10^6\K$, we can calculate the
mean density and hence the X-ray luminosity.

Both X-ray and other observations give estimates of the bubble radius.
Star counts (to a given limiting magnitude) and radio observations
give estimates of $L$, the bolometric stellar luminosity, or $Q$. The stellar
wind luminosity can then be estimated from either $L$ or $Q$, with the caveats
mentioned above. Finally, from the presence of Wolf-Rayet stars and/or
SN remnants, we have estimates of the cluster age. In terms of these
quantities, the X-ray luminosity is
\be  \label{eq:lx}
L_x\approx3\times10^{38}
\xi\left({L_w\over4\times10^{38}\ergs\s^{-1}}\right)^2
\left({20\pc\over r}\right)^3
\left({6\times10^6\K\over T}\right)^2
\left({3.6\times10^6\yr\over \tau}\right)^2,
\ee  
where we have assumed an X-ray cooling rate $\Lambda_x\approx 3\times
10^{-23}\xi\ergs\cm^3$, e.g., \citet{chu95}, $\xi$ is the metallicity relative
to solar, and $\tau$ is the age of the cluster. We have
scaled to values appropriate to Carina. As previously noted, this is
10,000 times too high.

In addition to Carina, we have calculated this predicted X-ray
luminosity for the LMC H II regions detected in X-rays by the {\em
  Einstein} satellite, and discussed in \citet{chu90}. We estimate the
bolometric luminosity of the H II regions in two ways. First, we use
the star counts from \citet{LH}; these authors report star counts for
stars with $m_V<14.7$, about $M_V=-3.85$ at the distance of the
LMC. Second, we use the observed radio free-free flux from
\citet{MBB6cm} to estimate $Q$ and thence $L$; the results from the
two methods are consistent, within a factor of $3$ or so. As in the
case of Milky Way bubbles, this indicates that the filling factor of
the H II gas is of order unity, and hence that the pressure in these
bubbles is not set by the X-ray emitting gas. From the bolometric
luminosity we then estimate $L_w$. Using these values,
eqn. (\ref{eq:lx}) over-predicts the X-ray luminosity, by factors
ranging from a few to five hundred.

Thus the picture for the LMC is consistent with our finding for
Carina.  Confusingly, other authors have found an {\em excess} of
observed X-ray emission compared to predictions of the Castor et
al. model \citep{chu90,chu95,WangHelfand91}. Why should this be?

\citet{chu90,chu95} and \citet{WangHelfand91} all use the scaling laws of Castor et al.~to predict the X-ray luminosity of bubbles in the LMC. 
The latter two give expression for $L_x$ that appear to contain only directly observed quantities, such as the bubble radius and the bubble expansion velocity. 
However, the coefficient relating $L_x$ to the observed quantities depends on the theoretical estimate for the bubble radius and central bubble density. 
This dependence results in an estimate for the stellar wind luminosity that is more than a factor of 10 smaller than the wind luminosity as estimated from either simple star counts, or from the free-free radio emission. 

In the case of \citet{chu90}, this
can be seen explicitly in their discussion of N51D, where they infer
$L_w=1.08\times10^{37}\ergs$ in their appendix B section IV. From the
star counts of \citet{LH} we estimate $L_w=2\times10^{38}\ergs$;
from the radio flux from \citet{MBB6cm} we find
$L_w=6\times10^{37}\ergs$. Using the lower value, and the observed
$R_b=48\pc$ eqn. (\ref{eq:lx}) yields $L_x=1.6\times10^{36}\ergs$,
about a factor $5$ larger than the observed $L_x=3.3\times10^{35}\ergs$.

Since their estimates of $L_x$ are smaller than the observed X-ray
luminosities, \citet{chu90} suggest that the X-ray emission of the LMC
objects is provided by supernovae that have gone off near the bubble
wall. We argue that, using the estimate given here, recourse to
supernovae is unnecessary. In addition, the morphology of the X-ray
emission in Carina does not support this picture, nor does that of the
objects studied by \citet{chu90} in the LMC. In Carina, as stated above,
the emission is slightly enhanced near the bubble walls, but not at a
particular point, as would be expected following a supernova.

Similarly, the X-ray images presented by \citet{townsley06} of N 157
(30 Doradus), especially their Figures 14 and 15, generally show
rather diffuse emission in the cavities surrounded by cold dusty gas,
reminiscent of what is seen in Carina. However, as \citet{townsley06} also remark,
in some cases the emission comes from the interior of the cavity,
while in others it comes from edge brightened regions with distinct
central voids, similar to Carina.  Two exceptions may be the
regions referred to as numbers 5 and 9, which \citet{townsley06}
suggest may be the result of supernovae interacting with the bubble
wall, a la \citet{chu90}. 

\subsection{Dynamics Controlled by H II Gas}

As we noted in section \S \ref{sec:Observations}, the H II gas pressure
in Carina, as estimated from the radio free-free emission, is similar
to the pressure in the X-ray emitting gas; both are similar to the
pressure where the outflows from photo-dissociation regions
shock. This was pointed out earlier by \citet{DM87}. Since the sound
speed of the cooler gas is comparable to the bubble expansion
velocities observed, this is not surprising. 

What is more surprising is that the free-free luminosity is consistent
with a filling factor for the H II gas that is of order unity (see eqn.
\ref{eqn:Q}) a fact apparently not noted before. This implies that the
gas pressure is set by the H II gas rather than the hot gas. It
suggests that the shocked stellar winds escape from the partially
confining bubble at the sound speed of the hot gas, as long as the
pressure in the hot gas exceeds the H II gas pressure in the
region. When enough hot gas has escaped that $P_x\approx P_{HII}$, the
escape will slow, as the hot gas is impeded by the cooler gas. This
sets up a rough pressure balance between the hot and H II gas.

\citet{Dorland86} and \citet{DM87} consider the loss of wind energy
(by an enhanced conduction by a non-Maxwellian tail of high energy
electrons, according to the latter). 
\citet{DM87} note that the thermal energy in hot gas in the Rosette as
well as the Carina nebula cavities is well below that input by stellar
winds over the age of the respective star clusters. These authors 
suggest that the excess energy is conducted away from the
hot gas to surrounding cold gas, rather than by simply flowing out through
holes in the bubble wall as suggested here and elsewhere. This would
alleviate the dynamical problem (the predicted bubble radius would be
smaller than in the Castor et al.~theory) and the X-ray luminosity
problem. 

In a similar vein, \citet{McKee84} consider the evolution of wind
blown bubbles in conjunction with H II regions. Like us, they conclude
that the pressure and hence the dynamics of the bubble are controlled
not by the $10^7\K$ gas, but by the H II gas. They suggest that the
energy density of the shocked wind is dissipated in situ, rather than
leaving the bubble. In the case of a high wind luminosity (as in
Carina), they suggest that the hot gas will engulf a large number of
globules, producing a large mass loading and resultant catastrophic
radiative cooling.

Examination of the Hubble Carina mosaic \citep{hubca}, which has a
spatial resolution of $\sim10^{-3}\pc$, does not reveal a large
population of evaporating globules in the interior of the bubble. If
the bubble wall consists of a large number of sub-resolution globules
it might be possible to dissipate the wind luminosity; the bulk of the
wind luminosity would be radiated near or just beyond the Lyman edge,
or possibly as optical light. Radiation beyond the Lyman edge would be
hard to detect due to absorption along the line of sight to Earth,
while optical emission would be hard to distinguish from emission
associated with ionizing radiation from the star clusters. However,
preliminary calculations of the emission region associated with an
evaporating globule suggest that a few percent of the luminosity would
emerge in the X-ray band; only if the globules are $\sim 100 $AU in
radius is the fraction of emission emerging in the X-ray band small
enough to escape detection. The conduction around such globules is
highly saturated; in fact, the mean free path for electrons to
thermalize is $\lambda\sim0.1\pc$. Hence, the globules would have to
be separated by many times their own radii to avoid an X-ray emission
fraction larger than observed. These small globules are surrounded by
a saturated conduction region roughly ten times their own radius; if
these saturated conduction regions overlap they will radiate X-rays
like a single, much larger globule. We defer further discussion of
this possibility to a later publication.

Neither \citet{Dorland86} nor \citet{McKee84} explain why
$P_{HII} \approx \sqrt{3Q/4\pi R_b^2\alpha_r}$ (where the latter assumes
a filling factor near unity). It is not clear why the
energy-loss rate of the hot gas would tune itself so that the filling
factor of both warm and hot gas is near unity. A large cooling luminosity
would lead to an essentially complete absence of hot gas, while a low
cooling luminosity would result in too high a hot gas pressure. Unlike
a simple mass loss scenario, an energy loss scenario requires some
fine tuning to match the observations.

Finally, we note that our leaky bubble models, which predict smaller
bubble radii and gas pressures, have implications for the prediction
of spectral energy distributions of starburst galaxies. As pointed out
by \citet{dopita05}, using a standard Castor et al.~bubble model
results in ionization parameters too small (by a factor of $10$ or
more) than that inferred from spectral energy distributions of such
galaxies. Both the smaller radii and the lower gas pressure (hence gas
density) will increase the ionization parameter associated with bubbles
surrounding massive star clusters.

\subsection{Ballistic Bubbles}

In the Castor et al.~scenario, if the shocked stellar wind gas from a
star or star cluster was confined to a bubble of volume $V$, the
pressure of the hot gas would be
\be  \label{eq:hotpressure} 
P_x= {2\over3}{5\over11}{L_w\tau\over V},
\ee  
where the energy in the bubble $E_b=(5/11)L_w\tau$ and $\tau$ is the
age of the star cluster. As already mentioned, this pressure is orders
of magnitude larger than the  observed pressure in
Carina, so the hot gas is not currently exerting the predicted
pressure on the surrounding dense gas. 

However, it is possible that in the past the hot gas was
confined. The force exerted by the gas on the bubble would have been
much larger than that exerted by the H II gas or by radiation,
resulting in a nearly impulsive expansion of the bubble. After this
early impulsive phase, the hot gas could then leak out of the bubble,
leaving the bubble to expand in a nearly ballistic manner. This could
happen, for example, if the density of the trapped hot gas grew large
enough that the radiative cooling time scale became shorter than the
expansion time scale \citep{shull}. We argue that this is not the case.

\begin{figure}
\begin{center}
\begin{tabular}{cc}
\epsscale{.50}
\plotone{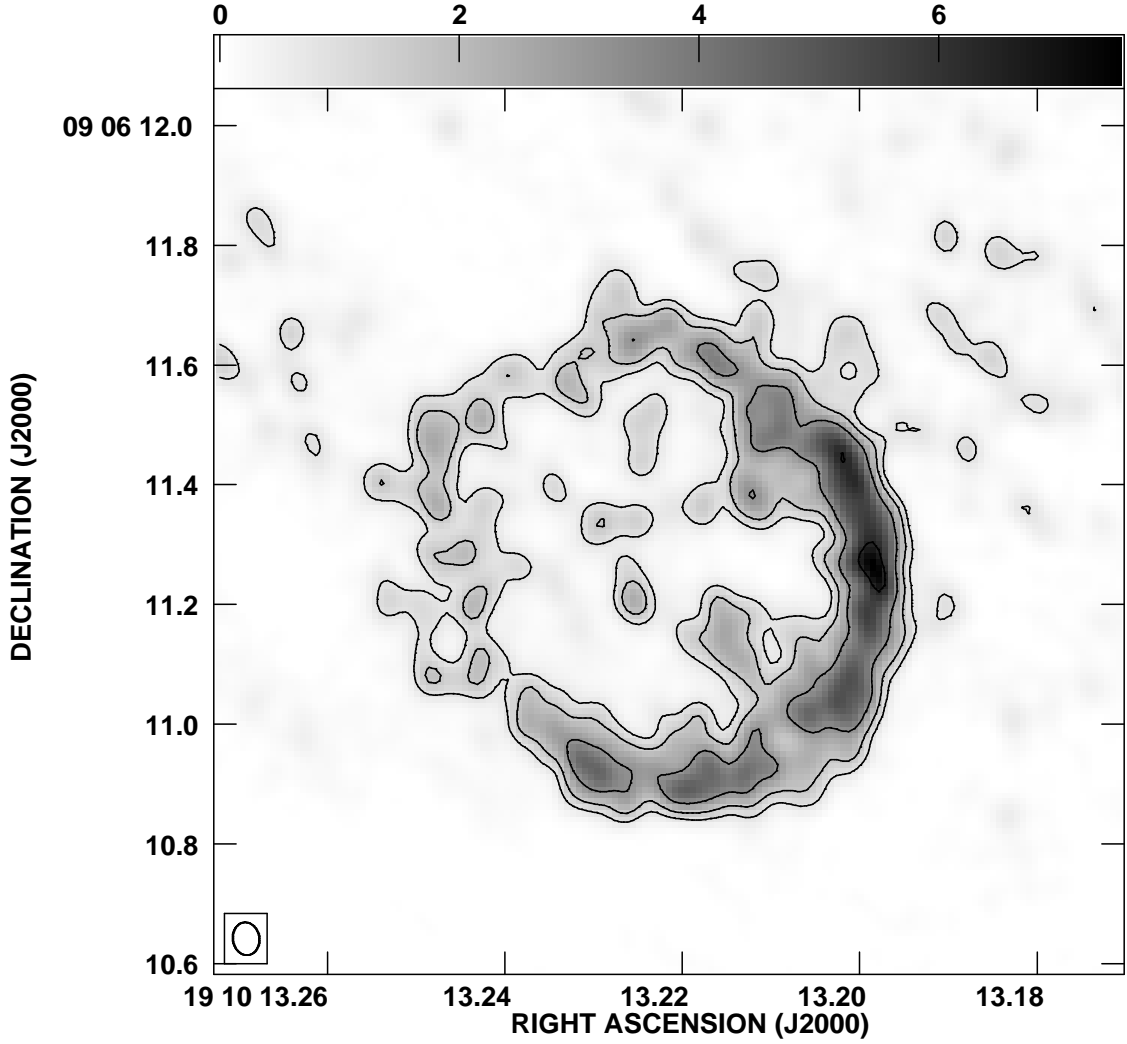}
\end{tabular}
\end{center}
\caption{{\small VLA 7mm observation of W49A/D from \citet{2005ApJ...624L.101D}. Beam size shown lower left is about 500 AU. The shell has a radius of $\sim 0.03 \pc$ and the column density through the shell is significantly variable.}}
\label{shell}
\end{figure}

First, high resolution radio images of ultracompact H II (UCHII) regions ($R\sim 0.1 \pc$)
by \citet{2005ApJ...624L.101D} show that most are either shell-like (see Fig.~\ref{shell}) or
cometary (a significant fraction are bipolar). The former two would
allow hot gas to leak out of the UCHII region. Observations show that the 
expansion velocities are typically $\sim 10 \kms$, e.g., \citep{Chakraborty97,Chakraborty99}.
The expanding shell will sweep up mass; if the motion is ballistic the velocity will decrease and the bubble will never reach sizes of order 10pc.

Second, there are UCHII regions that emit diffuse X-rays,
e.g., in the Welch ring of W49 \citep{tsujimoto} and in the Hourglass
region of the Lagoon nebula, M8 \citep{Rauw02}. In both cases the
X-ray luminosity is well below that expected under the assumption that
the shocked wind gas is confined in a bubble; it is also well below
$L_w$. We discuss each source in turn.

The W49 ultracompact ($r\sim0.15\pc$, $D=11.4\kpc$) source ${\rm G_X}$
has $L_x\approx3\times10^{33}$erg s$^{-1}$ and $T\approx7\keV$ \citep{tsujimoto},
much hotter than the plasma in Carina. The high temperature of the
X-ray gas is difficult to reconcile with the Castor et al.~model; such
hot gas is more likely to occur in a wind-wind collision. If this is
the case, then the luminosity of any truly diffuse, shocked, partially
confined gas is even smaller than the value given here.  The column
toward the source is $\Sigma\approx0.8\g\cm^{-2}$, suggesting a
dynamical pressure $P=\pi G\Sigma^2\approx 10^{-7}\dynes\cm^{-2}$, and
a mean cold gas density $\rho\approx2\times10^{-18}\g\cm^{-3}$.
This is consistent with their estimated hot gas pressure based on the emission
measure ($n_x\approx10\cm^{-3}$, $P_x\approx10^{-7}\dynes\cm^{-2}$). They suggest the driving
source is one or two O5 or O6 stars, with a luminosity in the range
$2-6\times10^5L_\odot$, consistent with the ionizing photon
luminosity $Q\approx10^{49}\s^{-1}$ for each of the five G sources
listed by \citet{depree97}. The corresponding wind luminosity is
$L_w\approx500L_\odot$.

The age of the source can be estimated in two ways. First, using the
mean density, the dynamical time is
$\tau_{dyn}=1/\sqrt{G\rho}\approx3\times10^{12}\s$. Second, using the
force exerted by the hot, X-ray emitting gas; as just noted, the
latter has a much lower density than the absorbing gas, but a similar
pressure. The outward acceleration is
$P_x/\Sigma\approx10^{-7}\cm\s^{-2}$. The dynamical time is then
$\tau_{dyn}\approx2\times10^{12}\s$, roughly the same as the estimate
based on the self-gravity of the cold gas.

Assuming that the stellar wind is trapped in a Castor style bubble,
the energy in the shocked gas is 
$E={5\over 11}L_w\tau\approx4\times10^{48}\ergs$, and the pressure
$P=2\times10^{-5}\dynes\cm^{-2}$, two orders of magnitude higher than
observed (and with an associated X-ray luminosity four orders of
magnitude too high). Since this pressure is so much higher than the
dynamical pressure, and the sound crossing time of the hot gas so much
smaller (by a factor of about 1000), the simplest interpretation is
simply that the shocked stellar wind gas escapes, as in Carina. The
radio images of the G sources in Figure 2 of \citet{2000ApJ...540..308D}
show a very incomplete ring, strengthening the argument for rapid escape.

The second case is the ultracompact H II region G 5.97 -1.17 associated
with the Hourglass nebulae in M8. It is believed to be powered by the
O7 V star Herschel 36; using values for $\dot M_w$ from \citet{rep04}
(not allowing for a clumped wind), we estimate
$L_w=2\times10^{36}\ergs$, while \citet{Rauw02} use a wind luminosity
a factor of $3$ smaller. They estimate a dynamical age of
$2\times10^4\yr$ based on the size of the Hourglass (they use
$R=0.2\pc$) and the observed expansion velocity of $10\kms$
\citep{Chakraborty97,Chakraborty99}. Using their values for $L_2$,
$\tau_{dyn}$, and $R_b$, we find a predicted hot gas pressure of
$P=4 \times 10^{-7}$ and $n_x \approx200\cm^{-3}$. The predicted
$L_x=7\times10^{35}\ergs$, larger than that observed by a factor of
$1000$. Using the full Castor et al.~model, \citet{Rauw02} note that
the predicted bubble radius is larger than that observed (but only by
a factor of 2) and that the predicted X-ray flux is
$L_x\approx10^{35}\ergs$. This smaller than our prediction by a factor of
$7$, consistent with their use of the (two times too large) Castor et
al. radius. Their predicted X-ray flux is still larger than observed
by a factor of $175$.

Radio observations by \citet{turner74} find a radius $r\sim 0.1\pc$
(half that suggested by Rauw et al.) and a flux at $11\cm$ of $0.54$
Jansky. The H II gas pressure is $1.5\times10^{-9}\dynes\cm^{-2}$ from
the observed radio flux, and $1.5\times10^{-8}\dynes\cm^{-2}$ using
$Q$ for an O 7 V star. This should be compared to the pressure of the
X-ray emitting gas $P_x=8\times10^{-9}\dynes\cm^{-2}$, and to the
radiation pressure $L/4\pi r^2c=1.6\times10^{-8}\dynes\cm^{-2}$.  All
these pressures are at least a factor $10$ below the pressure found
under the assumption that the wind luminosity is trapped in the
bubble. We conclude that, as for ${\rm G_x}$, the most likely
explanation of the low observed X-ray flux and the small bubble radius
is that the shocked stellar gas simply escapes from the bubble
interior.

\section{CONCLUSIONS}

We use both the Castor et al.~theory and the Chevalier and Clegg
theory for bubble evolution around massive star clusters to try to
understand observations of Milky Way and LMC bubbles.  The Castor
theory over-predicts X-ray luminosity in the Carina bubble by a factor
of 60 and expands too fast by a factor of 4.  In contrast, the
Chevalier and Clegg model under predicts the X-ray luminosity by a
factor of 20; since it is a steady state it has no expansion time.

These results suggest there is a partially confined hot, homogeneous
interior but there are also holes in the shell through which the hot
gas escapes like a free flowing wind.  We constructed a model in which
the confining shell covered only a fraction $C_f$ of the sky as seen
from the cluster, and allowed hot wind gas to escape at its sound
speed from the uncovered portion of the shell. As an example one model
consistent with observations of Carina is an isothermal sphere GMC
model with $R_G = 60 \pc$, $\alpha = 0.2$, and $C_f = 0.5$, $R_b =
16.5 \pc$, $L_x = 1.2 \times 10^{34} \ergs$, and $P_b = 1.1 \times
10^{-10}\,\dynes \cm^{-2}$.  This compromise is the best way to fit the many
observations of the Carina nebula and is likely the best solution for
all bubbles.  Our models suggest that the exact size, distribution and
evolution of holes does not matter much, it is the weighted average of
the fraction of total area that matters.

The bottom line is that the dynamics of X-ray bubbles are not strongly
affected by the presence of luminous stellar winds; rather, they are
controlled by a combination of the self-gravity of the surrounding GMC
gas, radiation pressure, and H II gas pressure. This conclusion is in
contrast to that of \citet{cas75} and \citet{shull}, but in agreement
with \citet{McKee84}. We argue that the X-ray fluxes of the bubbles
are {\em always} below those predicted by the Castor et al.~picture,
properly interpreted, i.e., using the observed rather than predicted
bubble radii. As noted by \citet{Dorland86,DM87}, the X-ray
observations show that the pressure in the hot gas is well below that
predicted by \citet{cas75}; unlike \citet{Dorland86,DM87}
and \citet{McKee84}, we argue that this is a result of leakage of hot
gas from the bubble, rather than leakage of energy in the form of
radiation. 

The results of the models in this paper are dependent upon the ISM
model used, both the run of mean density, and the fluctuations in the
surface density in different directions from the bubble
center. According to both simulations and observations, the latter is
distributed log-normally, meaning that in many directions the surface
density will be far below the mean surface density. We suggest that
hot gas escapes along these directions, which in our simplified models
we treat as holes in the bubble wall.

Full hydrodynamical models are needed to check that the
semi-analytical solutions shown in this paper are correct and to
understand both the early evolution of bubbles and the development of
holes in the shell.

\acknowledgments We would like to thank C. De Pree for allowing us to
use Fig. \ref{shell}. This research has made use of the SIMBAD
database, operated at CDS, Strasbourg, France, and of NASA's
Astrophysics Data System.  This work was supported by NSERC of
Canada. N.M. is supported in part by the Canada Research Chair
program, and by NSERC of Canada.

\appendix

\section{CHEVALIER \& CLEGG BUBBLES}

Chevalier \& Clegg assume a spherically symmetric wind with mass input
$\dot{M}_{w}$ and energy input $\dot{E} = L_w$.  For a smooth
transition from subsonic flow at the centre to supersonic flow at
large distances the Mach number must equal one at some radius, which
Chevalier \& Clegg take to be $r = R_{cl}$.

The flow is described by
\begin{equation}
\frac{1}{r^2}\frac{d}{dr} ( \rho u r^2 ) = \phi , \label{cc1}
\label{ccbeg}
\end{equation}
\begin{equation}
\rho u \frac{du}{dr} = -\frac{dP}{dr} - \phi u ,
\end{equation}
\begin{equation}
\frac{1}{r^2} \frac{d}{dr} \left[ \rho u r^2 
\left( \frac{1}{2} u^2 + \frac{5}{2} \frac{P}{\rho} \right) \right] = \Phi , \label{cc4}
\label{ccend}
\end{equation}
where $u$ is the wind velocity, $r$ is the radial co-ordinate, $\rho$
is the density, $P$ is the pressure, and $\gamma$, the adiabatic
index, $= 5/3$ throught this paper.  For $r < R_{cl} $, $\phi = \dot{M}_{w} / V$, $\Phi = \dot{E}
/ V$, $V = 4 \pi R_{cl}^3 /3$; for $r > R_{cl}$, $\phi = \Phi = 0$.
Equations (\ref{ccbeg})-(\ref{ccend}) have the solution
\begin{equation}
\left( \frac{5 + 1/\mathcal{M}^2}{6} 
\right)^{\left( -\frac{9}{14} \right) } 
\left( \frac{1+3/\mathcal{M}^2}{4} 
\right)^{\left( \frac{1}{7} \right) }
= \frac{r}{R_{cl}} ,
\end{equation}
for $r<R_{cl}$, and
\begin{equation}
\mathcal{M}^{3}
\left( \frac{1+3/\mathcal{M}^2}{4} 
\right)^{2}
= \left( \frac{r}{R_{cl}} \right)^2 ,
\end{equation}
for $r>R_{cl}$, where $\mathcal{M}$ is the Mach number.

The flow can be integrated in two sections, $r<R_{cl}$ and $r>R_{cl}$
using the outer boundary conditions of the first as the inner boundary
of the second.  Thus, it is possible to follow the evolution of the
wind and numerically integrate the expected X-ray luminosity from the
cluster edge out to arbitrarily large distances..

Initially a small radius ($r<<R_{cl}$) and small initial mach number,
$\mathcal{M}_0$ are assumed.  The true Mach number is then found
iteratively to 10 decimal places for that radius
\begin{equation}
\mathcal{M}_{i+1} = \left[ 6 
\left( \frac{1+3/\mathcal{M}^2}{4} 
\right)^{\frac{2}{9}} \left( \frac{r}{R_{cl}} 
\right)^{-\frac{11}{9}}-5 \right]^{-0.5} . \label{mach1}
\end{equation}
The calculated Mach number was then combined with the observed mass
loss rate and stellar wind luminosity for Tr 16 (Table 1) to find the
initial conditions (from \citet{che85})
\begin{eqnarray*}
\rho &=& 0.296  \dot{M}_{w}^{1.5}  L_w^{-0.5}  R_{cl}^{-2} , \\
n &=& \frac{\rho}{\mu} , \\
P &=& 0.118  \dot{M}_{w}^{0.5} L_w^{0.5} R_{cl}^{-2} , \\
T &=& \frac{P}{n  k_b} , \\
c_s &=& \sqrt{\frac{5  P}{3 \rho}} , \\
u &=& \mathcal{M}  c_s .
\end{eqnarray*}
From these initial conditions equations (\ref{ccbeg})-(\ref{ccend})
can be integrated to the cluster radius.  The integration continues
with the evolution equations (without the source terms outside the
cluster) and the appropriate Mach number iteration,
\begin{equation}
\mathcal{M}_{i+1} = \left( \frac{r}{R_{cl}} 
\right)^{\frac{2}{3}} \left(  \frac{4}{ 1 +
  3/\mathcal{M}_{i}}  
\right)^{\frac{1}{6} } . \label{mach2}
\end{equation}

\section{CASTOR et al.~BUBBLES}

For a bubble in the snow plough stage, $R_b$ is large enough that the
flow from c to b can be considered stationary and plane parallel.
\citet{wea77} considered the effects of relaxing this assumption and
found the results are only changed by $\leq$10\%, much less than the
observational errors for, e.g., the bubble radius in Carina.  For a
stationary plane parallel flow at constant pressure \citep{pen70}
\begin{equation}
\rho v \frac{dH}{dz} = \frac{d}{dz} \left[ K(T) \frac{dT}{dz} \right] - \rho \Lambda (T) ,
\label{tempint}
\end{equation}
where $T$ is temperature, $z = R_b - r$, $\rho v = \dot{M}_b / 4 \pi
R_b^2$, $M_b$ is the mass in region b, $H = 5 k_b T / 2 \mu$ the
specific enthalpy, $K(T) = C T^{5/2}$ is the thermal conductivity, and
$C = 1.2 \times 10^{-6}$ erg cm$^{-1}$ s$^{-1}$ K$^{7/2}$
\citep{spi62}.  Integrating equation(\ref{tempint}) from the b-c
boundary: $z=0$, $T \sim 0$ to the centre: $z=R_b$, $T=T_b$ gives the
mass evaporating from c to b as
\begin{equation}
\dot{M}_b = \frac{16 \pi}{25} \frac{\mu}{k_b} K(T) R_b .
\end{equation}
To get this solution we have assumed the cooling term, $\rho \Lambda
(T)$ is negligible in the transition zone between b and c.

Energy is constantly added into region b by the stellar winds with the
total luminosity $L_w$ and the dominant energy loss is through the
expansion of the bubble.  The energy in region b obeys
\begin{equation}
\dot{E_b} = L_w - 4 \pi R_b^2 P_b \dot{R}_b ,
\end{equation}
where $P_b$ is the pressure in region b and
\begin{equation}
\frac{4}{3} \pi R_b^3 P_b = \frac{2}{3} E_b .
\end{equation}

When considering radiative losses from region b it is necessary to
follow the evolution of region a (the hypersonic stellar wind region)
as eventually region b will cool enough to collapse.  Therefore, the
equations need to be altered to account for the fact that the size of
region a is not negligible.  In that case, the energy in region b
is given by
\begin{equation}
E_b = 2 \pi P_b (R_b^3 - R_a^3)  ,
\end{equation}
where $R_a$ is the radius of region a.
$R_a$ can be calculated from pressure balance at $R_a$.
The pressure from region b on a shell at $R_a$ is
\begin{equation}
P_b = \frac{1}{2 \pi} \frac{E_b}{R_b^3 - R_a^3}.
\end{equation}
The stellar wind luminosity is $L_w = \dot{M}_{w} v_w^2 /2$ so the
rate of momentum deposition by the stellar wind is
\begin{equation}
\dot{M}_{w} v_w = \frac{2 L_w}{v_w} .
\end{equation}
This then gives a pressure at radius $R_a$ of
\begin{equation}
P_{w} = \frac{2 L_w}{v_w} \frac{1}{4 \pi R_a^2} .
\end{equation}
Setting these pressures  equal gives
\begin{equation}
R_a = \left[    \frac{L_w}{v_w E_b}  (R_b^3 - R_a^3)  \right]^{1/2}  .
\end{equation}
Radiative losses within region b can be calculated by reference to the
cooling functions in \citet{sut93}
\begin{equation}
L_b \approx n_b^2 \frac{4}{3} \pi (R_b^3-R_a^3)  \Lambda (T)  .
\end{equation}

Including radiative losses into the Castor et al.~evolution equations
results in the following closed set of evolution equations
\begin{eqnarray}
\frac{d{\cal P}_c}{dt} & = & 4 \pi R_b^2 P_b \label{cas1} , \\
\frac{dR_b}{dt} & = & \frac{{\cal P}_c}{M_c}  , \\
\frac{dM_b}{dt} & = & C_1 T_b^{5/2} R_b^2 (R_b - R_a)^{-1} - 10^{-5} \frac{\mu}{k_b} L_b  ,\\
\frac{dE_b}{dt} & = & L_w - 4 \pi R_b^2 P_b \frac{dR_b}{dt} - L_b  ,
\label{cas4}
\end{eqnarray}
where ${\cal P}_c$ is the momentum of the shell and $C_1 = 4.13 \times
10^{-14}$ cgs.

The initial conditions of the bubble are taken from \citet{wea77} at
time $t_0 = 10^3$ years: {\setlength\arraycolsep{2pt}
\begin{eqnarray}
E_b &  = & \frac{5}{11} L_w t_0 , \label{casic1}\\
R_b & = & \left( \frac{125}{154 \pi} \frac{L_w}{\rho_0} \right)^{1/5} t_0^{3/5} ,\\
v_c & = &\frac{3}{5} \frac{R_b}{t_0} ,\\
V_b   & = & \frac{4}{3} \pi R_b^3,  \\
P_b & = & \frac{E_b}{V_b} ,  \\
M_b  & = & \frac{28}{205} \times 1.646^{5/2} \times V_b 
\frac{\mu}{k_b} \left( \frac{t_0 C}{R_b^2} \right)^{2/7} P_b^{5/7},   \\
\rho_b & = & \frac{M_b}{V_b} ,\\
n_b & = & \frac{\rho_b}{\mu}, \\
T_b & = & \frac{P_b}{n_b k_b} .\label{casic4}
\end {eqnarray}}
These give initial conditions close enough to the true values that the
numerical solution settles down by the cluster boundary and is a valid solution
for $R_b > R_{\hbox{\scriptsize{cl}}}$.

Initially the parameters were set up to match those used in
\citet{wea77} and results compared to Figure 6 of that paper as a check
of our code.  Once the code was tested parameters for Tr 16 could be
used (from Table 1).

\clearpage

\clearpage

\clearpage

\clearpage

\clearpage

\clearpage

\clearpage

\clearpage

\clearpage

\clearpage

\clearpage

\clearpage

\clearpage

\begin{table}
\begin{center}
\begin{tabular}{|l|l|}
\hline
Number of stars $\ge$ B2 & 72 \\ \hline
Total bolometric luminosity & $1.7\times 10^7$ L$_{\sun}$ \\ \hline
Total ionizing flux (Q) &  $5.9 \times 10^{50}$ s$^{-1}$ \\ \hline
Total mass loss rate $\dot{M}_{w}$& $1.1 \times 10^{-3}$ M$_{\sun}$/year  \\ \hline
Total stellar wind luminosity &  $6.7 \times 10^{4}$ L$_{\sun}$  \\ \hline
\end{tabular}
\end{center}
\caption{{\small Properties of Tr 16 in the Carina Nebula using all
    stars of spectral type B2 or earlier assuming standard mass loss
    rates \citep{rep04}. From \citet{smi06a}. } }\label{carina}
\end{table}

\begin{table}
\begin{center}
\begin{tabular}{|c|c|c|c|}
\hline
T/K & $L_d$ / erg s$^{-1}$ & $Q_e$ & $M_d$/$M_{\sun}$ \\ \hline
35 & $2.93 \times 10^{40}$ & $8.72 \times 10^{-5}$ & 6 610 \\ \hline
80 & $1.19 \times 10^{40}$ & $4.71 \times 10^{-4}$ & 18.2 \\ \hline
220 & $4.24 \times 10^{39}$ & $3.72 \times 10^{-3}$ & 0.014 \\ \hline
\end{tabular}
\end{center}
\caption{{\small Dust masses of the three components seen in \citet{smb07} with better dust modeling, compared to \citet{smb07} total dust mass of $9.6 \times 10^3 \hbox{M}_{\sun}$ } } \label{dustmass} 
\end{table}

\end{document}